\documentclass[
  prl,            
  aps,
  a4paper,
  reprint,
  twocolumn,      
  superscriptaddress, 
  nofootinbib,    
  aps,            
  floatfix        
]{revtex4-2}

\usepackage{cancel}
\usepackage{graphicx}   
\usepackage{amsmath}    
\usepackage{amssymb}    
\usepackage{bm}         
\usepackage{hyperref}   
\usepackage{tikz}
\usepackage{booktabs}
\usepackage[dvipsnames]{xcolor}
\usepackage{amsmath,amssymb,amsfonts}

\usepackage[normalem]{ulem}

\DeclareMathOperator{\arctanh}{arctanh}

\newcommand{\ie}{\textit{i.e.}, }
\newcommand{\eg}{\textit{e.g.}, }

\begin{document}
\title{Energy Transmission Across Holographic Conformal Interfaces in General Dimensions}

\author{Igal Arav}
\email{igal.arav@kuleuven.be}
\email{igala@hit.ac.il}
\affiliation{Instituut voor Theoretische Fysica, KU Leuven,
Celestijnenlaan 200D, B-3001 Leuven, Belgium}
\affiliation{Department of Physics, Faculty of Sciences, Holon Institute of Technology, 52 Golomb Street, Holon 5810201, Israel}

\author{Theodore Bertrand}
\email{bertrand@apc.in2p3.fr}
\affiliation{Université Paris Cité, CNRS, Astroparticule et Cosmologie, F-75013 Paris, France}

\author{Shira Chapman}
\email{schapman@bgu.ac.il}
\affiliation{Department of Physics, Ben-Gurion University of the Negev,
David Ben Gurion Boulevard 1, Beer Sheva 84105, Israel}

\author{Giuseppe Policastro}
\email{giuseppe.policastro@phys.ens.fr}
\affiliation{Laboratoire de Physique de l'\'Ecole Normale Supérieure, ENS, Université PSL, CNRS, Sorbonne
Université, Universit\'e de Paris, F-75005 Paris, France}

\author{Sebastian Waeber}

\email{waeber@post.bgu.ac.il}
\affiliation{Department of Physics, Ben-Gurion University of the Negev,
David Ben Gurion Boulevard 1, Beer Sheva 84105, Israel}

\date{\today} 

\begin{abstract}
Energy transport across conformal interfaces is universal in two spacetime dimensions, but its higher-dimensional counterpart has remained largely unexplored. In higher dimensions, scattering can depend on additional kinematic data, such as the angle of incidence, and conformal symmetry is far less restrictive. We use gravitational holography to study this problem. For a broad class of AdS$_d$-sliced domain-wall geometries, we derive a closed-form expression for the energy transmission coefficient in terms of the bulk warp factor. Our formula reproduces known two-dimensional holographic results and extends them to arbitrary dimension. Within this class, we  establish  different aspects 
of higher-dimensional universality. In particular, we show that the transmission is independent of the incidence angle and of the profile of the incident perturbation. We also derive bounds on transmission controlled by central-charge-like quantities characterizing the number of degrees of freedom of the theories on the two sides of the interface. This provides the first explicit result for reflection and transmission across conformal interfaces in higher dimensions, where no CFT-based results are currently available.
\end{abstract}


\maketitle 


\makeatletter
\def\@appendixcntformat#1{\csname the#1\endcsname}
\def\@hangfrom@appendix#1#2#3{%
  #1%
  \@if@empty{#2}{#3}{%
    #2\@if@empty{#3}{}{\enspace #3}%
  }%
}
\def\@seccntformat#1{\csname the#1\endcsname\quad}
\makeatother

\renewcommand{\thesubsection}{\thesection.\arabic{subsection}}
\renewcommand{\thesubsubsection}{\thesubsection.\alph{subsubsection}}

\renewcommand{\thesubsection}{\thesection.\arabic{subsection}}
\renewcommand{\thesubsubsection}{\thesubsection.\alph{subsubsection}}

\makeatletter
\renewcommand{\p@subsection}{}
\renewcommand{\p@subsubsection}{}
\makeatother

\setcounter{secnumdepth}{3}

\section{Introduction}
Defects and interfaces are useful probes of quantum field theories, appearing in settings from condensed-matter impurities and junctions to holographic domain walls and branes. A codimension-one defect, or interface, separates spacetime into two regions and can connect distinct quantum field theories or different phases of one theory, raising natural questions about how energy, charge, and information pass through it.

For conformal interfaces, residual conformal symmetry strongly constrains energy transmission. In two spacetime dimensions, these constraints are especially powerful, making energy scattering universal. The reflected and transmitted energy fractions are independent of the wave-packet shape or the local operator that creates it, and are fixed by the central charges together with a single coefficient, $c_{LR}$, appearing in the stress-tensor two-point function across the interface $\langle T_L T_R\rangle$ \cite{Quella:2006de,Meineri:2019ycm}. The relation of energy  and information transfer was  explored in \cite{Karch:2024udk}.

At strong coupling, however, the relevant field-theory coefficient is generally difficult to extract directly. Gravitational holography \cite{Ammon:2015wua,Aharony:1999ti} offers a complementary approach, representing defects, boundaries and interfaces geometrically as branes \cite{Karch:2000ct,Karch:2001cw,DeWolfe:2001pq,Bachas:2001hpy,Bachas:2001vj,Azeyanagi:2007qj,Takayanagi:2011zk,Fujita:2011fp} or domain-wall-like bulk solutions \cite{Bak:2003jk, Freedman:2003ax,Clark:2005te,DHoker:2006vfr,DHoker:2006qeo,DHoker:2007zhm,DHoker:2007hhe,Bak:2007jm,DHoker:2009lky,Chiodaroli:2009yw, Bak:2011ga, Gutperle:2012hy, Chen:2020efh,Bobev:2020fon,Arav:2020obl,Arav:2020asu}. In two dimensions, this has enabled explicit computations of the energy reflection and transmission coefficients for various interface models \cite{Bachas:2020yxv,Baig:2022cnb,Bachas:2022etu,Baig:2024hfc,Banerjee:2025agg,Chakraborty:2026wip}. Several such models have also been central to holographic studies of the information problem \cite{Almheiri:2019hni,Geng:2020fxl,Chen:2020uac,Chen:2020hmv,Hernandez:2020nem,Geng:2021hlu,Grimaldi:2022suv,Anous:2022wqh,Antonini:2025sur,Geng:2026asi}.

Much less is known about analogous energy-scattering questions in higher-dimensional interface CFTs. For $d>2$, the problem is richer, since energy transmission can depend on additional data such as the angle of incidence, while conformal symmetry is finite-dimensional and therefore less constraining than in two dimensions.

In this letter, we use holography to solve this energy-scattering problem explicitly for a broad class of strongly coupled models with gravitational duals. The result establishes aspects of higher-dimensional universality, including independence from the angle of incidence, 
and yields a closed-form expression for the energy transmission coefficient. Let us summarize the main results.

We consider $d$-dimensional ICFTs dual to holographic domain-wall geometries written in terms of AdS$_d$ slices,
\begin{equation}
ds_{(0)}^2 = dy^2 + a^2(y) ds_{\text{AdS}_d}^2,
\label{Janus-metricN}
\end{equation}
where $ds_{\text{AdS}_d}^2=\left(-dt^2+d\zeta^2+dx_i^2\right)/\zeta^2$ 
is the unit-radius AdS$_d$ metric and $y$ parametrizes the slicing direction. We require that the geometry is asymptotically AdS$_{d+1}$ near both boundaries, $y\to \pm\infty$, with the warp factor approaching that of empty AdS$_{d+1}$, 
$a(y)\sim a_{L,R}(y)\equiv l_{L,R}\cosh(y/l_{L,R})$, where $l_L$ and $l_R$ are the corresponding AdS$_{d+1}$ radii. 
Geometries of this form are known to be holographically dual to ICFTs
\cite{Gutperle:2012hy,Chen:2020efh}, including explicit
examples in different dimensions with identified dual theories and
embeddings in 10d or 11d supergravity \cite{Bak:2003jk,Clark:2004sb,Clark:2005te,DHoker:2006vfr,DHoker:2006qeo,DHoker:2007zhm,DHoker:2007hhe,Bak:2007jm,DHoker:2009lky,Chiodaroli:2009yw,Suh:2011xc,Bobev:2013yra,Bobev:2020fon,Arav:2020obl,Arav:2020asu}.

We find that the transmission coefficient for waves incident from the left (right) is
\begin{equation}
\mathcal{T}_{L,R} =
\frac{\int_{-\infty}^\infty a_{L,R}(y)^{-d}\,dy}
{\int_{-\infty}^\infty a(y)^{-d}\,dy}\,,
\end{equation}
where $a_{L,R}(y)$ is the warp factor of empty AdS$_{d+1}$ with radius $l_L$ ($l_R$). This expression reproduces  previously known results in $d=2$ \cite{Bachas:2020yxv,Baig:2022cnb,Bachas:2022etu} and extends them to general $d>2$. Moreover, in all cases studied, we prove the bound
\begin{equation}\label{eq:boundCT}
0\leq \mathcal{T}_L \leq \min \left(1,\frac{C^T_R}{C^T_L}\right),
\end{equation} 
where $C^T_{L,R}\propto l_{L,R}^{d-1}$ are the stress-tensor two-point function coefficients of the left and right vacuum theories \cite{Osborn:1993cr}.
Full transmission is not possible when $C_R^T<C_L^T$. 
In the holographic systems we study $C^T$ is also proportional to the usual measures of degrees of freedom, including the Weyl-anomaly coefficients in even dimensions \cite{Henningson:1998gx,Myers:2010tj}. The transmission bound can therefore be interpreted as reflecting the lack of a sufficient number of degrees of freedom available to carry the excitation across the interface.

Our results are universal in the sense that they are independent of the incidence angle, the detailed wave profile, and the particular source used to create the excitation within the class of models we study. 
The results of this Letter may be relevant to a broad range of problems, from strongly coupled interfaces and defects in condensed matter systems to questions in quantum gravity and holography.

\section{General Setup}
Consider a conformal interface separating two $d$-dimensional conformal field theories, CFT$_L$ and CFT$_R$, located at $x=0$, where $x$ is the coordinate normal to the interface. Such an interface breaks the $SO(d,2)$ symmetry group of the homogeneous CFTs into an $SO(d-1,2)$ subgroup, and in particular breaks the symmetry under shifts in the $x$ direction. Time shift symmetry, however, is preserved and thus energy is conserved.  
An important observable of such a system is the energy transmission coefficient, controlling the fraction of energy transmitted across the interface given an incident excitation propagating towards the interface from one side (which we take to be the left side).

We will focus on the total energy flux observable:\footnote{We are grateful to Lorenzo Bianchi for suggesting this observable as a probe of the energy flux during discussions at the Newton Institute program ``Quantum Field Theory with Boundaries, Impurities, and Defects''. This suggestion played an important role in shaping our analysis. We also thank him for sharing with us that related questions, including the CFT understanding of universality, are being pursued in ongoing work with collaborators \cite{Bianchi}.}
\begin{equation}\label{eq:energyfluxdef}
    \mathcal{E}(x) \equiv \int dt d^{d-2}x_i\, T_{tx}(t,x,x_i) \,,
\end{equation}
where $T_{\mu\nu}$ is the stress tensor and $x_i\, (i=1,\ldots,d-2)$ are the spatial coordinates tangent to the interface. Note that due to energy conservation, $\mathcal{E}(x)$ does not depend on $x$, nor on the side of the interface on which it is evaluated. It can thus be viewed as measuring the transmitted energy flux on the right, or alternatively the difference between the incident and reflected energy flux on the left. In the ICFT vacuum, we have $ \langle T_{\mu\nu} \rangle |_{\text{ICFT}}  = 0$ and therefore $\langle \mathcal{E} \rangle |_{\text{ICFT}} = 0$.

We will be interested in the expectation value $\langle \mathcal{E} \rangle |_{\text{ICFT}+\mathcal{J}_L}$ in the presence of small operator sources $\mathcal{J}_L$ on the left side of the interface, of order $ \mathcal{J}_L \sim O(\epsilon)$. These sources will include a small perturbation of the background metric $\delta g_{\mu\nu}$, which acts as a source for the stress tensor, but can also include sources for any other CFT operator. 
These sources can be viewed as creating some excited state on the left side and far away from the interface.

One can also relate them to a definition similar to the ones in \cite{Meineri:2019ycm,Hofman:2008ar}, which evaluate an ANEC observable, similar to $\mathcal{E}$, in a CFT state created by some smeared operator $\mathcal{O}_L$. In such a setup, the sources $\mathcal{J}_L$ could arise as sources for the various operators in the OPE of the two instances of $\mathcal{O}_L$ when calculating $\langle \mathcal{E} \rangle$ in this state. We will not pursue the details of this relation here, and simply settle for evaluating $\langle \mathcal{E} \rangle |_{\text{ICFT}+\mathcal{J}_L}$.\footnote{Note, however, that the ANEC based observables used in \cite{Meineri:2019ycm}  are evaluated in some arbitrary state but with no sources. Small sources, such as those introduced here, can be used as a technical tool to calculate the correlation functions involved in this evaluation, but, importantly, they do not imply that the results are only valid for a small incident excitation. A similar argument applies to the evaluation of $\langle\mathcal{E}\rangle$ in this paper.}

We will be interested in the leading $O(\epsilon)$ response of $\langle \mathcal{E} \rangle |_{\text{ICFT}+\mathcal{J}_L}$ in the presence of the sources. 
Although the sources can modify the local stress-tensor conservation, in the models we consider,  their $O(\epsilon)$ contribution in the time shift symmetric background is a total time derivative. Hence,
$\partial_x\langle\mathcal{E}(x)\rangle=O(\epsilon^2)$,
and we may treat the integrated flux as $x$-independent at order $O(\epsilon)$.

To define the transmission coefficient (from the left) $\mathcal{T}_L$, we divide this observable by $\langle \mathcal{E} \rangle |_{\text{CFT}_L+\mathcal{J}_L}$, \ie the expectation value of $\mathcal{E}$ in the homogeneous CFT$_L$ theory but with the same sources $\mathcal{J}_L$ turned on, and finally take the $\epsilon\to 0$ limit:
\begin{equation}\label{eq:transcoeff_def}
    \mathcal{T}_L \equiv \lim_{\epsilon\to0
    } \frac{\langle \mathcal{E} \rangle |_{\text{ICFT}+\mathcal{J}_L}}{\langle \mathcal{E} \rangle |_{\text{CFT}_L+\mathcal{J}_L}} \,.
\end{equation}
A priori, it may seem like this coefficient can depend on the sources $\mathcal{J}_L$. However, we will show that, at least in the context of the family of holographic models we consider, this is not the case and in that sense $\mathcal{T}_L$ is universal.\footnote{Note that for a smooth choice of sources with compact support, we expect $\langle \mathcal{E}\rangle$ to be finite, so that both the numerator and denominator in \eqref{eq:transcoeff_def} are finite. Since we later show that the transmission coefficient does not depend on the source, the ratio \eqref{eq:transcoeff_def} will continue to be finite even in the limit where the source has non-compact support. This is precisely the case in the calculation below.}

\section{Holographic Scattering}\label{sec:holosetup}

We will consider holographic models with actions consisting of an Einstein-Hilbert term, as well as matter fields which can include multiple minimally coupled scalar fields 
and a constant-tension thin-brane configuration. 
The action takes the form:
\begin{equation}\label{eq:general_action}
    S= \frac{1}{16 \pi G_N}\int d^{d+1}x \sqrt{-g} \left(R+\mathcal{L}_{\text{Matter}}\right),
\end{equation}
where $\mathcal{L}_{\text{Matter}}$ is the matter Lagrangian. 
We are interested in background solutions to the equations of motion with an AdS$_d$ sliced (Janus-like) metric of the form
\eqref{Janus-metricN}, with the asymptotic behavior specified there. 
The ansatz \eqref{Janus-metricN} respects the $SO(d-1,2)$ symmetry of the dual ICFT. The sources and VEVs corresponding to each bulk field are read off using the standard Fefferman-Graham (FG) prescription, after performing a diffeomorphism to a flat-sliced geometry near each boundary. From the dual CFT perspective this is equivalent to performing a Weyl transformation from an AdS to a flat background geometry.

In order to calculate the transmission coefficient \eqref{eq:transcoeff_def}, we perturb the metric and other fields around the solution \eqref{Janus-metricN} corresponding to the ICFT vacuum, and solve the linearized equations of motion for these perturbations. We will be working in the gauge  $\delta g_{y\mu}=0$.

The $t$ and $x_i$ shift symmetries of the background imply that the perturbations decompose in terms of plane waves in the $t$ and $x_i$ directions, with frequency $\omega$ and wave vector $k_i$. Since we are solving linear equations with this symmetry, the different modes will decouple. Therefore, the energy flux   $\mathcal{E}$ \eqref{eq:energyfluxdef} (which is the $\omega=k_i=0$ component of $T_{xt}$) will only depend on the $t$, $x_i$-independent perturbation modes (and sources).

At $\omega=k_i=0$ one can further decompose each perturbation mode in powers of $\zeta$, corresponding to different representations of $SO(d-1,2)$ (different defect operator dimensions). The different powers of $\zeta$ will therefore similarly decouple. 
Due to conservation, $\mathcal{E}$ is constant in the flat space theory, and thus in the AdS$_d$ Weyl frame it is expected to behave as $\zeta^{d-2}$. It will therefore only depend on modes that include metric sources of the form $\delta g_{\mu\nu} \sim \zeta^{d-2}$. Note that after performing a diffeomorphism to a flat-sliced geometry, the source and VEV have complementary exponents, summing to $d$. We indeed find a constant VEV and a source that behaves as $x^d$.\footnote{One might be worried that such a source is divergent as $x\to\infty$. However, a mode of this form can be thought of as the $\omega=0, x^d$ component in the decomposition of a general source that is well-behaved as $x\to\infty$.}

Finally, $\langle T_{tx} \rangle$ follows from the near boundary behavior of $\delta g_{tx}$. At $\omega=k_i=0$, symmetry dictates that the $\delta g_{tx}$ mode only couples to modes which are  invariant under the $SO(d-2)$ $x_i$ rotational symmetry (reflection symmetry for $d=3$) and odd under $t$ reflection. 
In our class of holographic models, this is the only such mode and it will therefore be sufficient to turn on a mode of the form $\delta g_{tx} \sim \zeta^{d-2}$, with all other field perturbations turned off. We demonstrate the above decoupling  explicitly in a specific example in \cite{SupMat}G.

Hence, we expect that in our holographic models, we can evaluate the  transmission  coefficient \eqref{eq:transcoeff_def} by turning on only one source $\delta g_{tx}^L \sim \beta_0^L \zeta^{d-2}$ for the metric on the left of the interface. The linear dependence on the coefficient $\beta_0^L$  will cancel between numerator and denominator of \eqref{eq:transcoeff_def}, leaving no source dependence.

This establishes universality in the context of our calculations. In particular, it means $\mathcal{T}_L$ does not depend on any sources which are not invariant under the $SO(d-2)$ rotation symmetry of the $x_i$ directions, implying there is no dependence on the angle of incidence of the excitation.

We will therefore proceed to perturb our metric as follows:
\begin{equation}\label{eq:bmodepert}
    ds^2 = ds_{(0)}^2+2\, \epsilon\, a^2(y)\beta(y) \zeta^{d-2} dt d\zeta.
\end{equation}
Under our assumptions about the matter content, the perturbation obeys the equation of motion (see \cite{SupMat}A)
\begin{equation}
\label{Eq:bN}
    d\, a'\,\beta'+a\, \beta'' =0,
\end{equation}
where primes denote derivatives with respect to $y$.

We solve equation \eqref{Eq:bN}  for $\beta$ 
with boundary conditions corresponding to a metric source on the left of the interface 
\begin{equation}\label{eq:asymsources}
       \lim_{y\to -\infty} \beta(y) = \beta_0^L, \qquad \lim_{y\to \infty} \beta(y) = 0.
\end{equation}
The solution reads
\begin{equation}\label{eq:betaBsol}
    \beta(y)= \tilde B \int_{y}^{\infty} a(y')^{-d} dy', \quad \tilde B\equiv\frac{\beta_0^L}{\int_{-\infty}^\infty a(y)^{-d}\,dy}.
   \end{equation}   
Since it doesn't matter which side we evaluate $T_{tx}$ on, we choose the right side, where there are no sources.\footnote{Notice that the conservation of $T_{tx}$ across the interface guarantees that even when evaluated on the left it would not receive anomalous contributions due to the presence of sources.
We demonstrate this explicitly in a specific example in \cite{SupMat}G. \label{FN: Ttx evaluation left}} 
The asymptotic behavior of $\beta(y)$ and 
$a(y)$ for $y \to \infty$ reads
\begin{equation}
\begin{split}
       a(y)& =\frac{l_R}{2}
       e^{y/l_R} +\mathcal{O}(1)\\
       \beta(y) &=\tilde B 
       \,\frac{2^d}{d\, l_R^{d-1}} e^{-d y/l_R} +\mathcal{O}(e^{-(d+1) y/l_R}).
    \end{split}
       \label{Eq:expansion}
   \end{equation}

We perform a coordinate change near the right boundary to FG coordinates with Minkowski boundary metric:
\begin{align}\label{eq:FGgeneric}
     y/l_R &= - \log \left(\frac{\rho}{2x}\right) +\dots,\quad 
     \zeta = x+\dots ,
 \end{align}
 where $\rho$ is the bulk radial coordinate and $x$ is the boundary direction orthogonal to the interface. 
 The dots represent subleading terms in $\rho$ that are needed to bring the metric to FG form. 
 After the change of coordinates, the metric  becomes 
 \begin{equation}
 \begin{split}
     ds^2 = ds^2_{(0)} + 2\,\epsilon \, \rho^{d-2} \frac{\tilde B }{dl_R^{d-3}} dt dx +\mathcal{O}(\rho^{d-1}).
\end{split}
\end{equation}
We therefore obtain for the stress tensor \cite{deHaro:2000vlm}: 
\begin{equation}\label{eq:TtxRcalc}
     \langle T_{tx}^R \rangle = \frac{\epsilon \beta_0^L}{16\pi G_N\int_{-\infty}^\infty a(y)^{-d}\,dy}.
 \end{equation}
A similar expression is obtained for the homogeneous CFT$_L$, finally leading to:\footnote{The analogous   representation of the transmission coefficient in $d=2$ in terms of the warp factor was also independently observed by A. Karch (private communication).} \footnote{Here, we used the definition  \eqref{eq:transcoeff_def} together with the fact that $\langle \mathcal{E}\rangle$ is
the $\omega=0$ component of $\langle T_{tx}\rangle$.} 
\begin{equation}
\label{eq:transmission_JanusN}
\mathcal{T}_L=\frac{\langle T_{tx}^R \rangle_{I}}{\langle T_{tx}^L \rangle_{0}} = \frac{\int_{-\infty}^\infty a_L(y)^{-d} dy}{\int_{-\infty}^\infty a(y)^{- d}\,dy}.
\end{equation}
One can show that this transmission coefficient obeys the bounds in equation \eqref{eq:boundCT}, see \cite{SupMat}B, similarly to the two-dimensional bound discussed in \cite{Meineri:2019ycm,Bachas:2020yxv,Bachas:2022etu}.\footnote{In fact, we show in \cite{SupMat}B that in some cases a  stronger bound can be proven.}

In the next two sections we illustrate our results in two simple examples: the thin-brane model and Janus geometries with a single scalar and flat potential.

\section{The Thin-Brane Model}\label{thin brane section} 
Our first example is the thin brane model, whose gravitational dual  consists of two AdS$_{d+1}$ patches with radii $l_L$ and $l_R$, respectively, separated by a thin brane with tension parameter $\Sigma$, anchored on the boundary at the interface location $x=0$. The gravitational action reads:
\begin{align}
\begin{split}
    & \hspace{-2pt}  S = S_{L}^{\text{bulk}}+S_{R}^{\text{bulk}}+S^{\text{brane}},\\
    & \hspace{-2pt} S_{L,R}^{\text{bulk}}\equiv\frac{1}{16 \pi G_N}\int d^{d+1}x_{L,R} \sqrt{-g} \left( R +\frac{d(d-1)}{l_{L,R}^2} \right)
    \\& \hspace{-2pt} 
    S^{\text{brane}}\equiv \frac{1}{8 \pi G_N}  \int_{\text{brane}} d^{d}x  \sqrt{-\gamma}(K_R-K_L-\Sigma) ,
    \label{Eq:action}
\end{split}
\end{align}
where $\gamma$ is the determinant of the induced metric and $K_{L,R}\equiv \gamma^{\alpha\beta}K_{\alpha\beta}$ are the traces of the left/right extrinsic curvatures \(K_{\alpha\beta}^{L,R}=\frac{1}{2}\mathcal{L}_n
\gamma_{\alpha\beta}\),  on the brane evaluated  using the normal
$n^\mu\partial_\mu=-\partial_y$ on both sides.

The equations of motion are Einstein's equations on each side together with 
 Israel's matching conditions on the brane \cite{Israel:1966rt}:
\begin{align}
   \label{Israel1}
    [\gamma_{\alpha\beta}]=0, \qquad 
    [K_{\alpha\beta}]-[K]\gamma_{\alpha\beta}&=\Sigma\gamma_{\alpha\beta},
\end{align}
where $[\cdot]\equiv\cdot_L-\cdot_R$ denotes the difference between the left and right sides.  
The equations of motion and matching conditions admit solutions with two asymptotically AdS regions when (see \cite{SupMat}C)
\begin{equation}
    \left|\frac{1}{l_R}-\frac{1}{l_L}\right|\leq\frac{\Sigma}{d-1}\leq\left(\frac{1}{l_R}+\frac{1}{l_L}\right).
    \label{boundN}
\end{equation}

The unperturbed interface solution, satisfying the Israel matching conditions on the brane, takes the form \eqref{Janus-metricN} with
\begin{equation}\label{eq:thinay}
    a(y) =
    \begin{cases}
    l_R \cosh\left(\frac{y-y_R^*}{l_R}\right) & y>0
    \\
    l_L \cosh\left(\frac{y+y_L^*}{l_L}\right) &y<0
    \end{cases}\,,
\end{equation}
where $y_*^{L,R}$ satisfy: 
\begin{equation}
\begin{split}\label{eq:thin-match123}
    &l_L \cosh\left(\frac{y_L^*}{l_L}\right)=l_R \cosh\left(\frac{y_R^*}{l_R}\right),\\
    &\frac{1}{l_L}\tanh \left(\frac{y_L^*}{l_L}\right)+
    \frac{1}{l_R}\tanh \left(\frac{y_R^*}{l_R}\right)
= \frac{\Sigma}{d-1} .
\end{split}
\end{equation}
The upper bounds on the tension \eqref{boundN} correspond to $y_{R,L}^{*}\rightarrow \infty$ while the lower bound corresponds to $y^*_{L,R}\rightarrow\pm \, \text{sign}(l_R-l_L)\times \infty$.

Substituting \eqref{eq:thinay} into \eqref{eq:transmission_JanusN} and changing variables yields:
\begin{equation}
\begin{split}\label{eq:thin_trans}
    &\mathcal{T}_L = \frac{l_L^{-(d-1)}\int_{-\infty}^{\infty} \frac{1}{\cosh^d \tilde y} d\tilde y }
    {l_L^{-(d-1)}\int_{-\infty}^{\frac{y^*_L}{l_L}} \frac{1}{\cosh^d \tilde y} d\tilde y +
    l_R^{-(d-1)}\int_{-\infty}^{\frac{y^*_R}{l_R}}\frac{1}{\cosh^d \tilde y} d\tilde y}
    \\
    &~~~=
    \frac{2 l_L^{-(d-1)}}
    {l_L^{-(d-1)}\tau\left[\frac{y^*_L}{l_L}\right]+l_R^{-(d-1)}\tau\left[\frac{y^*_R}{l_R}\right]
    },
\end{split}
\end{equation}
where we have defined
\begin{equation}
\begin{split}
\tau(\tilde y) \equiv
    1+\frac{2 \Gamma \left(\frac{d+1}{2}\right)
    \tanh \tilde y
    }{\sqrt{\pi } \Gamma \left(\frac{d}{2}\right)}  \, _2F_1\left(\frac{1}{2},1-\frac{d}{2};\frac{3}{2};\tanh ^2\tilde y
    \right).
\end{split}
\end{equation}

We can use the integral representation in the first line of \eqref{eq:thin_trans} together with the values of $y^*_L$ and $y^*_R$ corresponding to the tension bounds  \eqref{boundN} to show that
\begin{equation}\label{eq:thin_tension_bounds}
    \frac{C^T_R}{C^T_R+C^T_L}\leq \mathcal{T}_L\leq\min\left(1,\frac{C_R^T}{C^T_L}\right).
\end{equation}
We therefore see that zero transmission (total reflection) is not possible unless we set 
$C^T_R=0$. The upper bound does not allow for complete transmission when  $C_R^T<C_L^T$. Both conclusions are similar to those in the two-dimensional case \cite{Bachas:2020yxv}.

In the supplementary material, we rederive the thin-brane result directly from the Israel matching conditions \cite{SupMat}D, recast the derivation in  angular coordinates \cite{SupMat}E, and evaluate the transmission explicitly in several dimensions \cite{SupMat}F.

The result is  directly generalizable to the case with multiple branes, where the warp factor is split into more than two regions, as in \cite{Baig:2022cnb}. 
In this case, for general $d$, like in $d=2$, total reflection can be achieved. For example, for three regions, this requires a vanishing middle AdS radius $l_M \to 0$.\footnote{In this case, the denominator of \eqref{eq:thin_trans}, which becomes a sum of three different contributions, diverges even though the different integrals are finite. This 
limit requires infinite tension for both branes.}
However, unlike the $d=2$ case, the transmission is no longer generally a function of the sum of tensions.

In the specific case of $d=2$, \eqref{eq:thin_trans} simplifies: 
\begin{equation}
\mathcal{T}_{L}
=
\frac{2}{l_{L}}
\left[
\frac{1}{l_L}
+
\frac{1}{l_R}
+
\Sigma
\right]^{-1},
\end{equation}
matching the results of \cite{Bachas:2020yxv}.

\begin{figure*}[t]
    \centering
\includegraphics[width=0.595\linewidth,trim={1cm 1.1cm 0 0.8cm},
  clip]{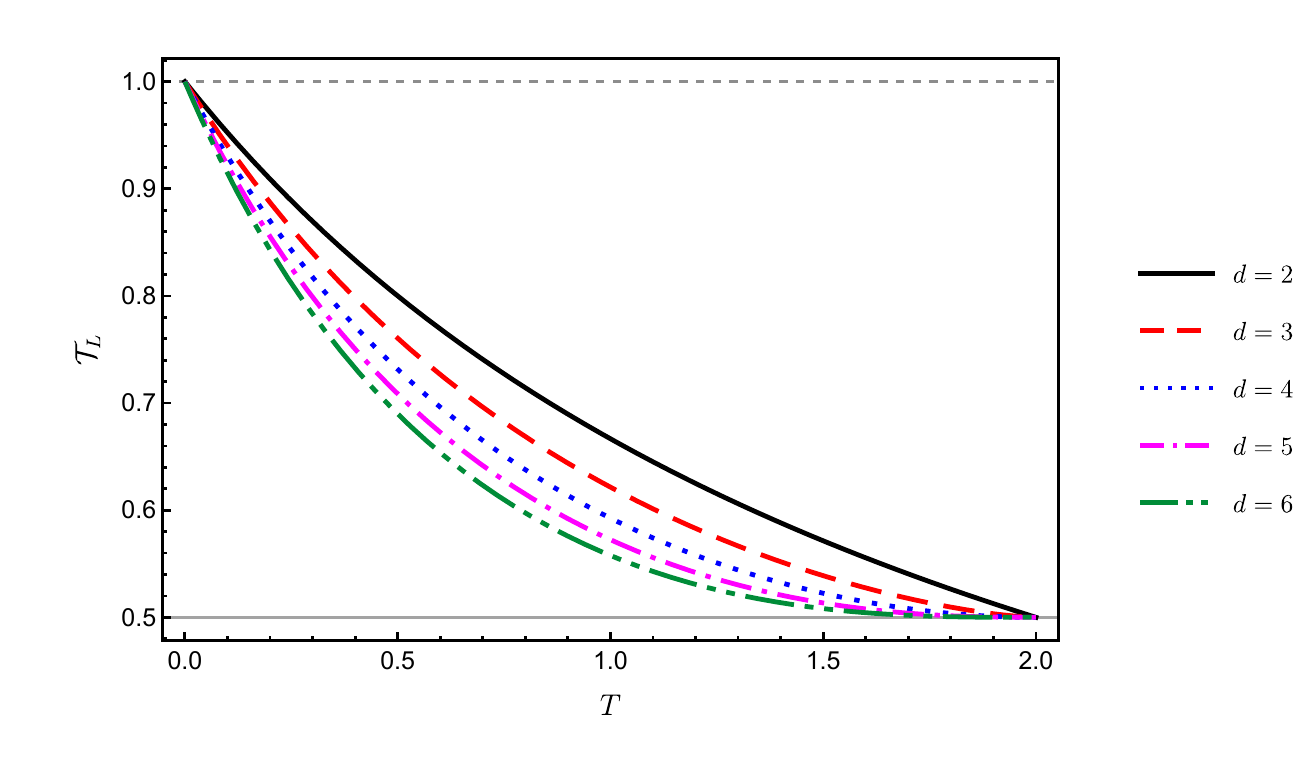}  ~~ \hfill
\includegraphics[width=0.376\linewidth]{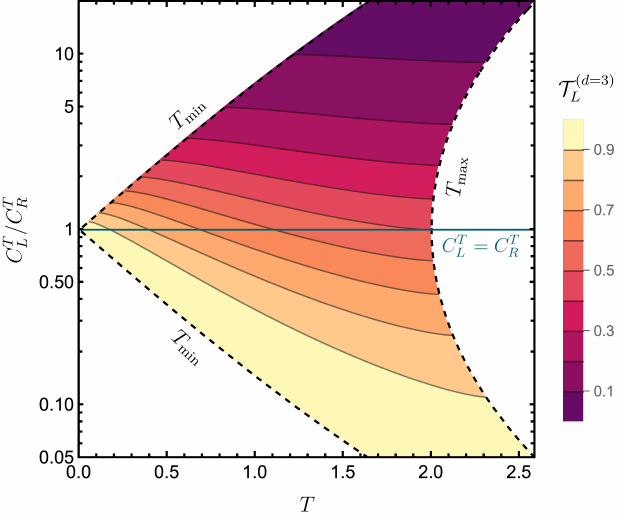} 
    \caption{The thin-brane transmission coefficient $\mathcal{T}_{L}$ as a function of the rescaled tension $T$, defined in equation \eqref{TLparams}.
Left: the symmetric case, $C_L^T=C_R^T$, for boundary dimensions $d=2,\ldots,6$. The upper (lower) transmission bound \eqref{eq:thin_tension_bounds} is indicated by the dashed (solid) horizontal gray  line. 
Right: the asymmetric $d=3$ case, shown as a contour plot in the $(T,C_L^T/C_R^T)$ plane.
The color scale indicates $\mathcal{T}_{L}^{(d=3)}$.
The black dashed curves labeled $T_{\min}$ and $T_{\max}$ denote the tension bounds  \eqref{boundN}, while the solid dark teal colored line marks the  $C_L^T=C_R^T$ slice. The plot illustrates that  complete transmission is not allowed when $C_R^T<C_L^T$, in agreement with the transmission bounds in equation \eqref{eq:thin_tension_bounds}. Further note that the tension bounds are symmetric under the exchange $C_L^T\leftrightarrow C_R^T$, as expected.}
    \label{fig:thin-brane}
\end{figure*}

To illustrate these results, in Fig.~\ref{fig:thin-brane} we plot the thin-brane transmission coefficient as a function of the dimensionless tension
\begin{equation}
    T=\frac{\Sigma\sqrt{l_L l_R}}{d-1}.
    \label{TLparams}
\end{equation}
The symmetric case, $C_L^T=C_R^T$, is shown in the left panel for several dimensions. The curves interpolate between unit transmission in the tensionless limit and $\mathcal{T}_L=1/2$ at the upper end of the allowed tension range, consistent with \eqref{eq:thin_tension_bounds}. In the right panel we show the asymmetric $d=3$ result as a contour plot in the $(T,C_L^T/C_R^T)$ plane. The admissible region is bounded by the tension bounds in equation \eqref{boundN}, and the   $C_L^T=C_R^T$ slice reproduces the corresponding $d=3$ curve in the left panel. Away from this slice, the lower and upper transmission bounds depend on the ratio $C_L^T/C_R^T$, in agreement with the bounds \eqref{eq:thin_tension_bounds}. Curves in other dimensions, both even and odd, are qualitatively very similar.

\section{Janus geometry example}\label{Janus section}
Our second example consists of Janus geometries described by the following action for a metric and a single scalar field:
\begin{equation}\label{Janus action}
    S= \frac{1}{16 \pi G_N}\int d^{d+1}x \sqrt{-g} \big(R-\partial^\mu \phi \partial_\mu \phi -2V(\phi) \big).
\end{equation}
Within the vacuum ICFT solution, the metric takes the form \eqref{Janus-metricN} 
and the scalar becomes $\phi=\phi(y)$.  
The scalar $\phi$ and the warp factor $a$ satisfy the equations of motion
\begin{equation}
\begin{split}
\label{eq:Janus}
\frac{d(d-1)}{2}
\left[
\left(\frac{a'}{a}\right)^2 + \frac{1}{a^2}
\right]
& =
\frac{1}{2}(\phi')^2 - V(\phi),
\\
 \phi'' + d \phi' \frac{a'}{a} - \frac{dV}{d\phi} & = 0 \, .
\end{split}
\end{equation} 
The transmission coefficient  is then simply given by substituting the warp factor solution into  \eqref{eq:transmission_JanusN}. 
In two dimensions the
result reproduces the results of \cite{Bachas:2022etu}.\footnote{For the $d=2$ Janus case, the source-based evaluation of the transmission coefficient is demonstrated  explicitly in \cite{SupMat}G. We devote a separate subsection of \cite{SupMat}G to spelling out intermediate steps previously omitted in the scattering prescription used in earlier literature \cite{Bachas:2022etu}.}

Let us now specialize to the case of a flat potential $V(\phi)=-\frac{d(d-1)}{2 l^2}$, corresponding to identical asymptotic radii $l_R=l_L =l$. 
The second equation of \eqref{eq:Janus} is solved by \cite{Freedman:2003ax}\footnote{Note that $b$ here differs from the parameter $b$ used in \cite{Bachas:2022etu} for the $d=2$ case.  The two are related by $b_{\text{here}}= b_{\text{there}}(2-b_{\text{there}})/4$.\label{foot:bbhatconventions}}
\begin{equation}
    \phi=\phi_0+l^{d-1}\sqrt{b \,d(d-1)}\int_0^y \frac{1}{a(y)^d} 
    d\tilde y,
    \label{phi_function}
\end{equation}  
where \(b\) and \(\phi_0\) are dimensionless constants. 
Note that although we use $b$ to
conveniently parametrize this family of solutions, the defect 
is more directly characterized by the asymptotic dilaton jump
$\Delta\phi
    \equiv
    \phi_\infty-\phi_{-\infty}$. 
This quantity encodes the jump in the marginal coupling associated with the operator dual to the dilaton across the
interface. 
Comparing this expression with
\eqref{eq:transmission_JanusN} gives a relation of the form $\Delta\phi
    \sim \sqrt{b}/\mathcal{T}_L(b)$.

Substituting \eqref{phi_function} into the first equation in
\eqref{eq:Janus}, we obtain the warp factor by inverting
\begin{equation}
    y/l= \int_{a_0}^{a(y)}\frac{da}{a\sqrt{1-\left(\frac{l}{a}\right)^2+b\left(\frac{ l}{a}\right)^{2d}}},
    \label{A_function}
\end{equation}
 where $a_0$ is the largest positive root of the denominator of \eqref{A_function}, at which the integrand diverges. 
 The parameter $b$ should be in the range
\begin{equation}
    0\leq b \leq b_{\text{max}}= \frac{1}{d} \Big(\frac{d-1}{d}\Big)^{d-1}.
    \label{b_max}
\end{equation}
Using \eqref{A_function} and 
\eqref{eq:transmission_JanusN}, the transmission coefficient can be expressed as 
 \begin{equation}
 \mathcal{T}_L 
=
\frac{\sqrt{\pi}\,\Gamma\!\left(\frac{d}{2}\right)}
{2\,\Gamma\!\left(\frac{d+1}{2}\right)}
\left[
\int_{\alpha_0}^{\infty}
\frac{\alpha^{-d}\,d\alpha}
{\sqrt{\alpha^2-1+b \alpha^{2-2d}}}
\right]^{-1},\label{eq:trans_int_jan}
 \end{equation}
  where we have defined the dimensionless warp factor $\alpha=a/l$ and $\alpha_0$ is again the denominator's largest positive root. 
At $b=0$, $\Delta\phi=0$, $\alpha_0=1$, and $\mathcal{T}_L=1$, as expected in the absence of a defect. Conversely, as $b\to b_{\max}^{-}$, the root becomes double and the integral diverges logarithmically, yielding $\Delta\phi\to\infty$ and $\mathcal{T}_L\to0$, \ie complete reflection.

\begin{figure}[t]
    \includegraphics[width=1\linewidth,trim={1.4cm 1.3cm 0 0.8cm},
  clip]{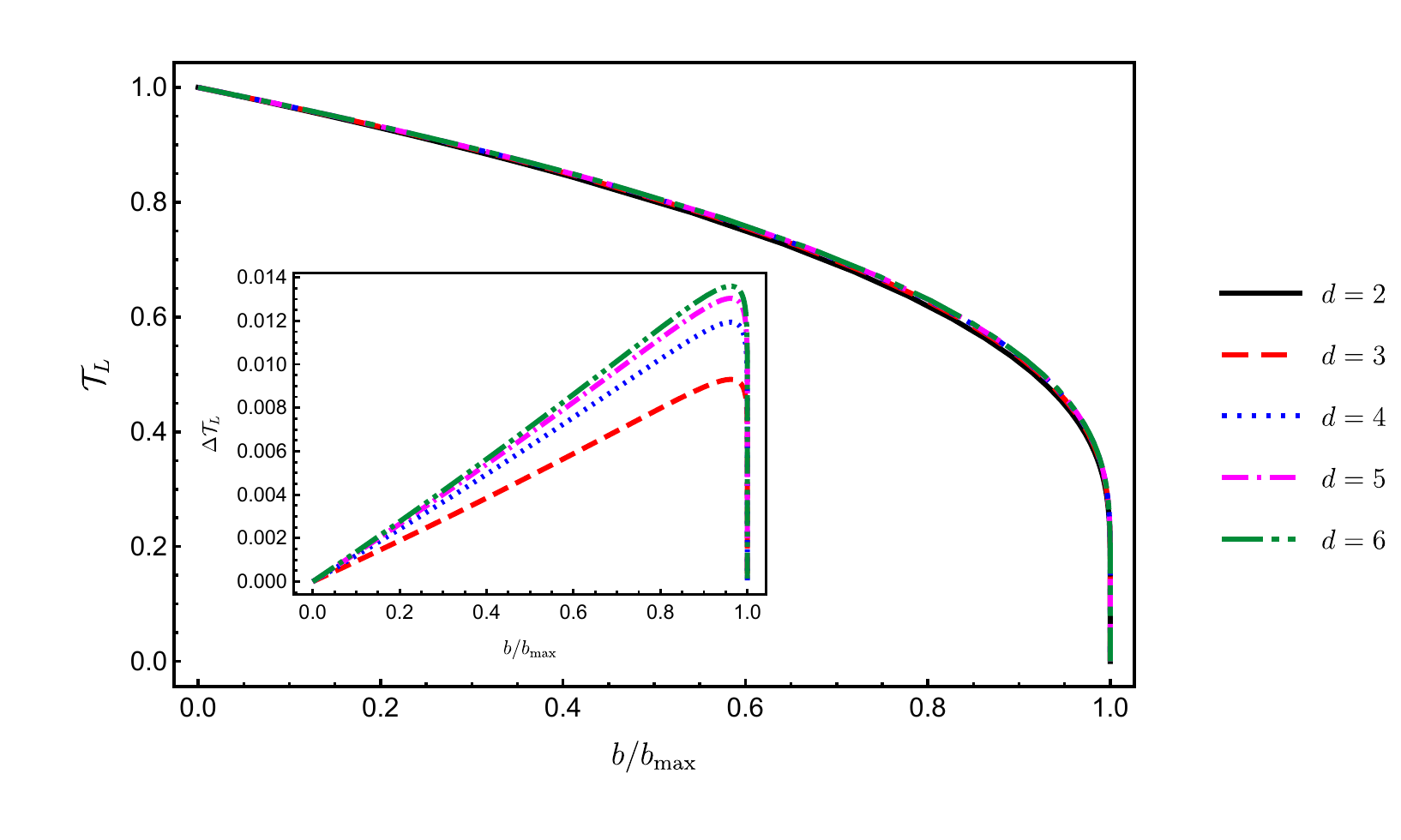}
  \caption{The transmission coefficient \(\mathcal{T}_L\) for Janus
geometries with flat potential as a function of
\(b/b_{\mathrm{max}}\) for \(d=2,\ldots,6\). The inset resolves the weak
dimensional dependence by showing
\(\Delta\mathcal{T}_L\equiv\mathcal{T}_L-\mathcal{T}_L^{(d=2)}\), the
difference relative to the \(d=2\) result.}
\label{fig:transmission_Janus}
\end{figure}

The results are shown in Fig.~\ref{fig:transmission_Janus} for several
dimensions as functions of \(b/b_{\mathrm{max}}\). Note that as
\(b\to b_{\mathrm{max}}^{-}\), the numerical evaluation becomes delicate.\footnote{This is because the two positive roots of the expression under the square root in \eqref{eq:trans_int_jan} coalesce at $\alpha_c=\sqrt{(d-1)/d}$. We isolate the resulting singularity by adding and subtracting a term proportional to $[\alpha\sqrt{\alpha-\alpha_0}\sqrt{\alpha+\alpha_0-2\alpha_c}]^{-1}$, which we integrate analytically, leaving a regular remainder for numerical integration.} Consistent with the limits above, the
transmission coefficient interpolates from \(1\) to \(0\) in every dimension. In
contrast to the thin-brane model, the curves for different dimensions
are nearly indistinguishable as a function of $b/b_{\max}$. Their differences reveal a weak
dimensional dependence, with the transmission increasing slightly with
$d$.\footnote{Note that these observations might change if one chooses a different parameterization of the family of defects, rather than $b/b_{\max}$.}

\section{Discussion}

We have studied energy transmission in holographic ICFTs in general dimension $d>2$. For a broad class of domain-wall geometries, we derived a universal expression for the transmission coefficient, independent of the incidence angle and the details of the source representing the perturbation profile. This points to a higher-dimensional analogue of the universality familiar from two-dimensional conformal interfaces \cite{Meineri:2019ycm}.  We establish this  universality only within the class of models and perturbations considered here. In particular, our analysis of classical Einstein gravity theories with minimally coupled scalars or thin-branes does not extend to other types of bulk matter or to general operator deformations beyond the single-trace sector described by these classical bulk fields.

An important question is therefore how far this universality extends. Holographically, it would be useful to test the result in more general matter-coupled backgrounds and in top-down supergravity solutions, or to include quantum corrections. It would further be interesting to explore universality with respect to more general definitions of the transmission coefficient on the gravity side directly. This could include  going beyond the linearized order in the perturbation studied in this paper and determining whether the results are consistent with the linearized analysis, as implied by CFT arguments. If indeed universality extends beyond the class of models studied here, a direct CFT derivation would be particularly valuable in identifying the principles and CFT data responsible for it. 

Our source-based construction provides a starting point for a broader holographic dictionary for defect and interface observables. Such a dictionary could shed light on the bulk origin of the
agreement in $d=2$ between the present prescription and the
scattering setup of Refs.~\cite{Bachas:2020yxv,Bachas:2022etu} and in particular on the role of the additional no-outgoing-wave
condition required there, but absent here. 
It would also be interesting to follow our construction
through dimensional reduction to a three-dimensional
Chamblin--Reall setup, similar to the one in Ref.~\cite{Ge:2026elr}, which studies a distinct bulk scattering experiment.

Finally, it would be interesting to connect these results to microscopic condensed-matter realizations. For example, while certain pinning defects were argued to factorize in \cite{Popov:2025cha}, such behavior is not expected to be generic. Explicit QFT examples or lattice models could provide concrete tests of how generic our observations regarding the transmission are.

\begin{acknowledgments}

\section*{Acknowledgments}
We would like to thank Costas Bachas, Lorenzo Bianchi, Andreas Karch, Marco Meineri and Francesco Nitti for useful discussions. 
SC, TB and GP would like to thank the Isaac Newton Institute for Mathematical Sciences, Cambridge, for support and hospitality during the program ``Quantum Field Theory with Boundaries, Impurities, and Defects'' where part of this work was undertaken. GP would also like to thank the Galileo Galilei Institute in Florence for hospitality during the workshop ``Defects and Extended Excitations in Quantum Field Theory, Quantum Matter and Statistical Models''. The program was supported by EPSRC grant no EP/R014604/1. IA is supported by FWO projects G094523N and G003523N, as well as by KU Leuven grant C16/25/010. The work of SC and SW is supported by the Israel Science Foundation (grant No. 1417/21), by the German Research
Foundation through a German-Israeli Project Cooperation (DIP) grant “Holography and
the Swampland”, by Carole and Marcus Weinstein through the BGU Presidential Faculty
Recruitment Fund, by the ISF Center of Excellence for theoretical high-energy physics, by
the VATAT Research Hub in the Field of quantum computing and by the ERC Starting
Grant dSHologQI (project number 101117338). The work of SW is further supported by the Kreitman fellowship
program at Ben-Gurion University.

\end{acknowledgments}

\bibliography{MyBibBMode}

\makeatletter
\def\@appendixcntformat#1{\csname the#1\endcsname}
\def\@hangfrom@appendix#1#2#3{%
  #1%
  \@if@empty{#2}{#3}{%
    #2\@if@empty{#3}{}{\enspace #3}%
  }%
}
\def\@seccntformat#1{\csname the#1\endcsname\quad}
\makeatother

\appendix
\renewcommand{\thesubsection}{\thesection.\arabic{subsection}}
\renewcommand{\thesubsubsection}{\thesubsection.\alph{subsubsection}}
\setcounter{secnumdepth}{3}



\onecolumngrid   
\newpage         

\makeatletter
\onecolumn@grid@setup
\let\set@footnotewidth\set@footnotewidth@one
\let\compose@footnotes\compose@footnotes@one
\global\count\footins=1000\relax
\makeatother

\setcounter{page}{1}




\begin{center}
{\LARGE Supplemental Material}
\end{center}

\section{Derivation of the perturbation equation of motion}

In this section, we derive the linearized equation of motion \eqref{Eq:bN} for the metric perturbation \eqref{eq:bmodepert} around an AdS$_d$ sliced background geometry  \eqref{Janus-metricN} for an action of the form \eqref{eq:general_action}. We also comment on some of the features of this equation. We assume the matter fields include multiple minimally coupled scalar fields $\phi^I$ with a Lagrangian of the general form:
\begin{equation}\label{eq:GeneralMatterLagrangian}
    \mathcal{L}_\text{Matter} =  - g^{\mu\nu} G_{IJ}(\phi)\partial_\mu \phi^I \partial_\nu \phi^J - 2 V(\phi) \, ,
\end{equation}
where $G_{IJ}(\phi)$ is a metric on some target space parametrized by the scalar fields $\phi^I$ and $V(\phi)$ is a general potential. The matter can also include a thin brane configuration, with each brane having an action of the form $S^{\text{brane}}$, see  equation \eqref{Eq:action}.\footnote{The results also extend to certain cases with brane localized scalar fields, as we explain below.} 
The equation \eqref{Eq:bN} has two interesting features: First, it has no terms proportional to $\beta(y)$, and is therefore invariant under constant shifts $\beta \to \beta + \text{const}$. Second, it has no explicit dependence on the background matter fields. While this equation can be derived in a straightforward way starting from the full action, we would like to highlight a shorter derivation that makes these features manifest.

We start by recalling, as explained in section \ref{sec:holosetup}, that the background geometry and fields are invariant under both time reversal $t \to -t$ and $SO(d-2)$ rotations of $x_i$ (or reflection $x_i \to -x_i$). Since the perturbation of the form \eqref{eq:bmodepert} is the only ($\omega=k_i=0$) mode which is odd under time reversal but $SO(d-2)$ invariant, it naturally decouples from all other perturbation modes (of both the metric and the matter fields / branes), and it is therefore consistent to set all other modes to vanish.

Next, we consider applying a small diffeomorphism of the form $ t \to t - \epsilon\, C \frac{\zeta^{d+1}}{d+1}$, where $C$ is some constant, to the background metric and fields. They transform as follows:
\begin{equation}
    ds_{(0)}^2 \to ds_{(0)}^2 + 2 \epsilon\, C a^2(y) \zeta^{d-2} dt d\zeta \,, 
    \qquad \phi^I(y)\to \phi^I(y) \,.
\end{equation}
When thin branes are present, the fluctuation of each brane is likewise unchanged by this diffeomorphism because its unperturbed embedding does not depend on $t$.
Therefore we see that performing such a diffeomorphism is equivalent to shifting the perturbation by $\beta(y) \to \beta(y) + C$. Since the equations of motion should be invariant under a diffeomorphism, we conclude that they must be invariant under constant shifts of $\beta$. In other words, if we denote the Einstein equations by:\footnote{Here, $T_{\mu\nu}^{\text{Matter}}$ 
 is defined by $T_{\mu\nu}^{\text{Matter}}\equiv
-\frac{2}{\sqrt{-g}}\,
\frac{\delta}
{\delta g^{\mu\nu}}
\left(\int \sqrt{-g}\,\mathcal{L}_{\mathrm{Matter}}\right)
$ and includes the bulk cosmological constant. Localized brane contributions are treated separately later through the Israel matching conditions.}
\begin{equation}
    E_{\mu\nu} = R_{\mu\nu} - \frac{1}{2} R g_{\mu\nu} -\frac{1}{2} T^\text{Matter}_{\mu\nu} =0,
\end{equation}
then the \emph{linearized} equations of motion around the background $\delta E_{\mu\nu}(\beta'',\beta',\beta)$  must satisfy $\frac{\partial E_{\mu\nu}}{\partial\beta} =0$, and therefore one can write them as:
\begin{equation}
    \delta_{\beta'',\beta'} E_{\mu\nu}  = 0 \,,
\end{equation}
where we have denoted
\begin{equation}
    \delta_{\beta'',\beta'} \mathcal{F}(\beta'',\beta',\beta) \equiv \left.\frac{\partial \mathcal{F}}{\partial\beta''}\right|_{\beta''=\beta'=\beta=0} \beta'' +  \left.\frac{\partial \mathcal{F}}{\partial\beta'}\right|_{\beta''=\beta'=\beta=0} \beta' \, .
\end{equation}
However, for bulk matter corresponding to \eqref{eq:GeneralMatterLagrangian} the stress tensor does not depend on derivatives of the metric and thus 
\begin{equation}
    \delta_{\beta'',\beta'}\, T^\text{Matter}_{\mu\nu}= 0 \,.
\end{equation}
Furthermore, since the Ricci scalar is time reversal invariant (as are the background fields), we have $\delta R=0$ and therefore $ \delta_{\beta'',\beta'}(g_{\mu\nu} R ) = 0$.
We are therefore left with the equation:
\begin{equation}
    \delta_{\beta'',\beta'} \, R_{\mu\nu} = 0 \,.
\end{equation}
We see that indeed the equation does not depend on the matter sector, and evaluating this equation we find:
\begin{equation}
     \delta_{\beta'',\beta'} \, R_{\zeta t} = -\frac{1}{2} \epsilon \zeta^{d-2} a \left[a \beta'' + d a' \beta'\right]  = 0 \,.
\end{equation}

In the presence of thin branes, we can make a similar argument for the matching condition \eqref{Israel1} around each brane. If we denote the (second) matching condition by:\footnote{Note that the first matching condition implies that $\beta$ is continuous at the brane location and therefore on the brane $\beta_L=\beta_R\equiv \beta$.}
\begin{equation}
    I_{\alpha\beta} \equiv [K_{\alpha\beta}]-[K]\gamma_{\alpha\beta}-\Sigma\gamma_{\alpha\beta} \,,
\end{equation}
the linearized condition $\delta I_{\alpha\beta}(\beta'_L,\beta'_R,\beta)  $ must satisfy $ \frac{\partial I_{\alpha\beta}}{\partial \beta} = 0$, and therefore it can be written:
\begin{equation}
    \delta_{\beta'_L,\beta'_R} I_{\alpha\beta} = 0 \,.
\end{equation}
On the other hand, due to time reversal invariance we must have $\delta[K]=0$, and therefore 
\begin{equation}
     \delta_{\beta'_L,\beta'_R} I_{\alpha\beta} =  \delta_{\beta'_L,\beta'_R} [K_{\alpha\beta}] = 0 \,,
\end{equation}
and evaluating this we obtain: 
\begin{equation}
    \delta_{\beta'_L,\beta'_R} [K_{\zeta t}] = -\frac{1}{2} \epsilon\zeta^{d-2} a_B^2 (\beta'_L - \beta'_R) = 0 \,,
\end{equation}
where $a_B$ is the warp factor  $a(y)$ evaluated at the location of the thin brane.
Therefore the matching condition simply implies $\beta'$ is continuous across the brane. 
Note that this argument is still valid when the brane action depends on contributions either from the bulk scalars or from additional scalar degrees of freedom localized on the brane itself, as long as the scalars don't couple to the (extrinsic or intrinsic) curvature of the brane, since their contribution to the matching condition will not depend on the derivative of the metric.

Let us now show an alternative derivation of the same equation. 
We start from the observation that for an action of the form \eqref{eq:general_action} with \eqref{eq:GeneralMatterLagrangian}, if $U^\mu$ is a Killing vector of the metric solution to the EOMs and $\mathcal{L}_U \phi^I =0 $, then:\footnote{This  follows from the Killing vector equation and from a well-known identity for Killing vectors \cite[Eq.~(C.3.9)]{Wald:1984rg} which together give
$\nabla_\mu Q^{\mu\nu}=\nabla_\mu\nabla^\mu U^\nu=-R^\nu{}_{\rho}U^\rho$.
Then, contracting the Einstein equation with $U^\rho$ and using
$\mathcal{L}_U\phi^I=0$ gives
$R^\nu{}_{\rho}U^\rho=\tfrac12\mathcal{L}U^\nu$. Combining the two observations yields the relation \eqref{eq:KillingVectorRelation}.}
\begin{equation}\label{eq:KillingVectorRelation}
    \nabla_\mu Q^{\mu\nu} = -\frac{1}{2} \mathcal{L}\, U^\nu \,,
\end{equation}
where $  Q_{\mu\nu} = \nabla_{[\mu}U_{\nu]} $ and $ \mathcal{L} = R + \mathcal{L}_\text{Matter}$. For our solutions, $U=\frac{\partial}{\partial t}$ is a Killing vector, both with and without the perturbation, and thus \eqref{eq:KillingVectorRelation} is satisfied for any $\beta(y)$ satisfying the equations of motion. Taking the linearized variation of this equation in the presence of the perturbation we obtain: 
\begin{equation}
    \delta \left[ \nabla_\mu Q^{\mu\nu} \right]= \nabla_\mu \delta Q^{\mu\nu} = -\frac{1}{2} (\delta\mathcal{L}) \, U^\nu = 0 \,,
\end{equation}
where we have used the identity $\nabla_\mu Q^{\mu\nu}=(-g)^{-1/2}\partial_\mu\!\left[(-g)^{1/2}Q^{\mu\nu}\right]$ for antisymmetric tensors together with 
the fact that $\delta(\sqrt{-g}) = 0$ and $\delta\mathcal{L} = 0$ (which follow from time reversal invariance), as well as $\delta U^{\mu}=0$ (by definition). 
We again see that the dependence on the matter sector drops, and evaluating the $\zeta$ component of this equation we obtain: 
\begin{equation}
    \nabla_\mu \delta Q^{\mu \zeta} = a^{-d} \partial_y\left( a^d \frac{1}{2}\epsilon\zeta^d \beta' \right) = \frac{\epsilon\zeta^d}{2a} \left( a \beta'' + d a' \beta' \right) = 0 \,.
\end{equation}

In the presence of thin branes, we can use a similar relation.  For each brane the following holds:
\begin{equation}
    n_\mu [Q^{\mu\nu}] = [K^\nu_\mu] U^\mu = - \mathcal{L}_\text{brane} U^\nu \,,
\end{equation}
where $\mathcal{L}_\text{brane} = K_R - K_L - \Sigma$, $n_\mu$ is the normal to the brane, 
and we have used the matching condition \eqref{Israel1} together with the fact that $U^{\nu}$ is a Killing vector tangential to the brane. Taking the linearized variation of this equation we obtain:
\begin{equation}
    \delta\left(  n_\mu [Q^{\mu\nu}] \right) = n_\mu [\delta Q^{\mu\nu}] = - (\delta\mathcal{L}_\text{brane}) U^\nu =0 \,,
\end{equation}
where we have used the fact that $\delta n_\mu=0$ and $\delta\mathcal{L}_\text{brane}=0$ (which follow from time reversal invariance). Evaluating the $\zeta$ component of this equation we obtain:
\begin{equation}
     n_\mu [\delta Q^{\mu\zeta}] = \frac{1}{2}\epsilon \zeta^d (\beta'_L-\beta'_R) = 0 \,.
\end{equation}

\section{Proof of bounds on transmission}
\label{proof_of_bounds}

We would like to prove the bound \eqref{eq:boundCT} on the transmission coefficient given in the expression \eqref{eq:transmission_JanusN}. We assume a regular solution to the EOMs corresponding to the action \eqref{eq:general_action}, with a metric of the form \eqref{Janus-metricN} and with asymptotics $a(y)\sim a_{R,L}(y)\equiv l_{R,L}\cosh(y/l_{R,L})$ as $y\to\pm\infty$. We also assume the bulk matter stress tensor $T^\text{Matter}_{\mu\nu}$ satisfies the null energy condition. 
We start by restricting to the case with no thin branes, such that the solution is smooth for $-\infty<y<\infty$, and we comment at the end on adapting the proof in the presence of thin branes as well.

In the background corresponding to the ICFT vacuum, the matter stress tensor is invariant under the $SO(d-1,2)$ symmetry of the background metric, and we can therefore in general parametrize it as follows 
\begin{equation}\label{eq:proofboundstresstensor}
    \frac{1}{2} T_{\mu\nu}^\text{Matter} dx^\mu dx^\nu = \chi(y) dy^2 + \left[ \gamma(y) - \frac{1}{2} \chi(y) \right] ds^2_{(0)}  = 
    \left( \frac{1}{2} \chi(y) + \gamma(y) \right) dy^2 + a^2(y) \left(\gamma(y) - \frac{1}{2} \chi(y)\right) ds_{\text{AdS}_d}^2 \,,
\end{equation}
where $\gamma(y),\chi(y)$ 
 are some scalar functions.
In order for this to be consistent with the aforementioned asymptotics, the matter stress tensor should asymptote to one corresponding to the correct cosmological constant on either side of the geometry and thus we must have:
\begin{equation}\label{eq:gammadeltaasympt}
\lim_{y\to-\infty} \gamma(y) = \gamma_L \equiv \frac{d(d-1)}{2} \frac{1}{l_L^2},\qquad
\lim_{y\to\infty} \gamma(y) = \gamma_R \equiv \frac{d(d-1)}{2} \frac{1}{l_R^2},\qquad
\lim_{y\to\pm\infty} \chi(y) = 0 \,.
\end{equation}
Note that for the case of a single scalar field (as in \eqref{Janus action}) these parameters are:
\begin{equation}
    \gamma(y) = -V(\phi(y)), \qquad \chi(y) = (\phi'(y))^2 \,. 
\end{equation}
Given this parametrization, the EOMs can be written as follows: 
\begin{equation}\label{eq:EOMforProof}
\left(\frac{a'}{a}\right)^2+\frac{1}{a^2} = f(y),
\qquad \frac{a''}{a} = g(y),
\end{equation}
where
\begin{equation}\label{eq:proofboundfmgm}
    f=f_m \equiv\frac{2}{d(d-1)}\left( \frac{1}{2} \chi + \gamma\right),
    \qquad
    g=g_m \equiv -\frac{1}{d} \chi + \frac{2}{d(d-1)}\gamma \,.
\end{equation}

The conservation of the bulk stress tensor in the $y$ direction gives the following relation:
\begin{equation}\label{eq:fmderfromstresstensorcons}
    \frac{d(d-1)}{2} \frac{df_m}{dy} = -d \frac{a'}{a} \chi \,. 
\end{equation}
Finally, from the null energy condition, with null vector 
$u^\mu\partial_\mu\equiv \partial_t\pm \frac{a(y)}{\zeta}\partial_y$ in the $y-t$ plane, we obtain: 
\begin{equation}\label{eq:deltaposfromNEC}
   \frac{1}{2} T^{\text{Matter}}_{\mu\nu} u^\mu u^\nu = \frac{a^2}{\zeta^2}\chi \geq 0,
\end{equation}
and therefore $\chi(y) \geq 0$.

\medskip

In order to prove the bound on the transmission coefficient, we will make use of the following lemma.
\medskip

\noindent
\textbf{Lemma.}
Suppose we have a solution $a(y)$ to \eqref{eq:EOMforProof} with $f=f_m$ and $g=g_m$, and additionally a different solution $a_0(y)$ to the same equations but with $f=f_0(y)$ and $g=g_0(y)$
such that $a'(0)=a_0'(0)=0$, 
and suppose the following three conditions are satisfied:
\begin{enumerate}
\item[(a)] $a(y)>0$, $a_0(y)>0$ for all $-\infty<y<\infty$,
\item[(b)] $g(y)\leq g_0(y)$, and $g_0(y)> 0$ for all $-\infty<y<\infty$,
\item[(c)] $f(0)\geq f_0(0)$. 
\end{enumerate}
Then $a(y)\leq a_0(y)$ for all $-\infty<y<\infty$.

\medskip

\noindent
\textbf{Proof of Lemma.}
Since $a'(0)=0$, from \eqref{eq:EOMforProof} we have 
\begin{equation}    
\frac{1}{a^2(0)}=f(0),
\qquad
\frac{1}{a_0^2(0)}=f_0(0).
\end{equation}
Condition (c) then gives
\begin{equation}\label{eq:condia0ga}
f(0)\geq f_0(0)
\quad \Rightarrow \quad
a(0)\leq a_0(0).
\end{equation}
Combined with the fact that $a'(0)=a_0'(0)=0$ and conditions (a) and (b), a simple comparison theorem then implies that $a(y)\leq a_0(y)$.\footnote{Indeed, define $r(y)\equiv a_0(y)/a(y)$, which is positive by
condition~(a). Using Eq.~\eqref{eq:EOMforProof}, we find $\bigl(a^2 r'\bigr)'=a^2(g_0-g)r\geq0$.
Moreover, $r'(0)=0$ and \eqref{eq:condia0ga} gives
$r(0)=a_0(0)/a(0)\geq1$. It follows that $r'\geq0$ for $y>0$
and $r'\leq0$ for $y<0$, and hence $r(y)\geq r(0)\geq1$ on
both sides. Therefore $a(y)\leq a_0(y)$ for all $y$.} 
This concludes the proof of the Lemma.

\medskip

Now consider a solution to the equations of motion with the assumed asymptotics \eqref{eq:gammadeltaasympt}. The function $\gamma(y)$ must have some finite global maximum $\gamma_\text{max} = \max_{-\infty\leq y \leq\infty} \gamma(y)$, and either it reaches it at some finite value $y_0$, \ie  $\gamma(y_0) = \gamma_\text{max}$, or $\gamma_\text{max} = \gamma_{L/R} $ in which case we will denote $y_0=\mp \infty$, respectively. 
This implies
\begin{equation}\label{eq:gammamaxbound}
    \gamma(y) \leq \gamma(y_0) = \gamma_\text{max} \quad \text{for all } y\,.
\end{equation}

Next, consider the function $f_m(y)$. The equation \eqref{eq:fmderfromstresstensorcons} along with the inequality \eqref{eq:deltaposfromNEC} implies that $\operatorname{sign}(f'(y))=-\operatorname{sign}(a'(y))$. Therefore
\begin{equation}
\begin{split}
\label{eq:proofboundsfinftybehavior}
&y\to -\infty:
\quad
a'\to -\infty
\quad \Rightarrow \quad
f'(-\infty)\geq0 \,,\\
& y\to +\infty:
\quad
a'\to +\infty
\quad \Rightarrow \quad
f'(\infty)\leq0.
\end{split}
\end{equation}
We can then conclude that there must be some finite point $y_\ast$ such that $f(y_\ast)$ is a global maximum, \ie
\begin{equation}\label{eq:fmaxbound}
f(y_\ast)\geq f(y)
\quad
\text{for all } y.
\end{equation}
Moreover, due to equation \eqref{eq:fmderfromstresstensorcons}, this point must be a turning point for $a(y)$, namely $a'(y_\ast)=0$.\footnote{Notice that this is true even if $\chi$ has an isolated zero at $y=y_\ast$ as can be seen by examining the behavior of $a$ and $f_m$ in a neighborhood of $y_\ast$.} 
We can use the shift symmetry in $y$ of the EOMs to choose $y_\ast =0$.

Let us denote by $f_0,g_0$ the constant functions:
\begin{equation}
f_0(y)=g_0(y)=\frac{2}{d(d-1)}\gamma_\text{max},
\end{equation}
and suppose $a_0(y)$ is a solution to the equations \eqref{eq:EOMforProof} with $f=f_0$ and $g=g_0$, with the boundary condition $a_0'(0)=0$. Note that $a_0(y)$ simply corresponds to an empty AdS$_{d+1}$ solution with a radius $l_0$ satisfying
\begin{equation}
    \frac{1}{l_0^2} = \frac{2}{d(d-1)} \gamma_\text{max} \,,
\end{equation}
namely $a_0(y)=l_0 \cosh(y/l_0)$. Since $\gamma_\text{max}\geq \max(\gamma_L,\gamma_R)$,  this implies:
\begin{equation}\label{eq:l0leqlLlR}
    l_0 \leq \min(l_L,l_R) \,.
\end{equation}

Let us now show that the solutions $a(y)$ and $a_0(y)$ satisfy the conditions of our Lemma. First, by our choice of $y=0$ and boundary condition for $a_0(y)$ we have $a'(0)=a_0'(0)=0$. Condition (a) is satisfied due to the assumed regularity of the solutions. 
For condition (b), we have from \eqref{eq:deltaposfromNEC} and \eqref{eq:gammamaxbound}
\begin{equation}
    g(y) = -\frac{1}{d} \chi + \frac{2}{d(d-1)} \gamma \leq \frac{2}{d(d-1)} \gamma_\text{max} = g_0(y),
\end{equation}
as well as
\begin{equation}
    g_0(y)= \frac{2}{d(d-1)}\gamma_\text{max} \geq \frac{2}{d(d-1)} \gamma_{L/R} > 0 \,.
\end{equation}
Finally, for condition (c) we have from \eqref{eq:fmaxbound} and \eqref{eq:deltaposfromNEC}:
\begin{equation}
    f(0) \geq f(y_0) = \frac{2}{d(d-1)}\left( \frac{1}{2} \chi(y_0) + \gamma(y_0) \right) \geq \frac{2}{d(d-1)} \gamma(y_0) = \frac{2}{d(d-1)} \gamma_\text{max} = f_0(0) \,.
\end{equation}
Therefore, from the lemma, we conclude $a(y)\leq a_0(y)$.
Using the result \eqref{eq:transmission_JanusN} and \eqref{eq:l0leqlLlR}, this immediately implies the bound for the transmission coefficient as:

\begin{equation}
\mathcal{T}_L = \frac{\int_{-\infty}^\infty a_L(y)^{-d} dy}{\int_{-\infty}^\infty a(y)^{- d}\,dy} \leq \frac{\int_{-\infty}^\infty a_L(y)^{-d} dy}{\int_{-\infty}^\infty a_0(y)^{- d}\,dy} = \frac{l_L^{-(d-1)}}{l_0^{-(d-1)}}\leq \min\left(1,\frac{l_R^{d-1}}{l_L^{d-1}} \right)\,.
\end{equation}
Note that when $l_0<\min(l_L,l_R)$  (equivalently when $\gamma_{\max}>\max(\gamma_L,\gamma_R)$), the $l_0$-dependent bound appearing in the middle step of the above inequality is strictly stronger than the general bound presented in the final step. For a single scalar, this occurs when $V(\phi)$ falls below both of its asymptotic values somewhere in the bulk in which case $l_0^2=-d(d-1)/(2V_{\min})$ where $V_{\min} = \min_{-\infty<y<\infty} V(\phi(y))$.

Let us now comment on the case where thin branes are also present. For this case, we consider the distributional contribution of the branes to the stress tensor separately from that of the bulk matter as given in \eqref{eq:proofboundstresstensor}, so that the functions $\gamma, \chi$ defined above are piecewise smooth. In the presence of the branes, the function $f_m$ given in equation \eqref{eq:proofboundfmgm} does not get any new localized contributions, whereas $g_m$ gets a contribution as follows (see also equation \eqref{eq:thin-background-delta}): 
\begin{equation}
    g_m \equiv -\frac{1}{d} \chi + \frac{2}{d(d-1)}\gamma + g_{\text{branes}}\,,
    \qquad
    g_{\text{branes}} = -\sum_{i=1}^n \frac{\Sigma_i}{d-1} \,\delta(y-\bar{y}_i) \,,
\end{equation}
where $\bar{y}_i$ are the positions of the branes and $\Sigma_i>0$ (which follows from the null energy condition).\footnote{Note that here $\Sigma_i$ is the on-shell tension of the brane, which could contain the contribution of on-shell brane scalars.}

In order to account for the branes, we first note that the Lemma given above remains valid when the function $g(y)$ has distributional contributions of the form given above (while keeping $g_0(y)$ smooth and regular), as long as $\bar{y}_i\neq 0$ and $\Sigma_i>0$.\footnote{Recall that $y=0$ is not an arbitrary point. The Lemma requires it to be an extremum of $a(y)$. As we show in a moment the possibility of a brane located there is ruled out.} Since $\gamma(y)$ is piecewise smooth, it still has a global maximum $\gamma_\text{max}$ at some point $y_0$. The main adjustment required is with  the choice of the point $y_\ast$. 

The function $f(y)$ is piecewise smooth and still satisfies \eqref{eq:proofboundsfinftybehavior}, and so there is still some finite point $y_\ast$ where $f(y_\ast)$ is a global maximum. Suppose this maximum is reached in the interval $[\bar{y}_i,\bar{y}_{i+1}]$ (where here $ 0\leq i\leq n $, $\bar{y}_0=-\infty$ and $\bar{y}_{n+1}=\infty$). If $\bar{y}_i<y_\ast<\bar{y}_{i+1}$ and some bulk matter is present in this interval (so that $\chi(y)$ doesn't identically vanish in the interval), 
then, by shift symmetry, we can choose $y_\ast=0$, the Lemma holds and the rest of the proof follows. 
Suppose then that the maximum of $f(y)$ is reached as $y\to \bar{y}_i^+$ (for $i=1,\ldots,n$). Then the jump in $f(y)$ over the brane must satisfy $\Delta f = f_R - f_L \geq 0$. The matching condition at the brane gives (see \eqref{eq:thin-background-israelN}):
\begin{equation}
    \frac{a_L'(\bar{y}_i)}{a(\bar{y}_i)} = \frac{a_R'(\bar{y}_i)}{a(\bar{y}_i)} + \frac{\Sigma_i}{d-1} \,,
\end{equation}
and therefore:
\begin{equation}
    \Delta f=  \left(\frac{a_R'(\bar{y}_i)}{a(\bar{y}_i)}\right)^2 - \left(\frac{a_L'(\bar{y}_i)} {a(\bar{y}_i)}\right)^2 = -\frac{\Sigma_i}{d-1} \left( 2\frac{a_R'(\bar{y}_i)}{a(\bar{y}_i)} + \frac{\Sigma_i}{d-1} \right) < -2 \frac{\Sigma_i}{d-1} \frac{a_R'(\bar{y}_i)}{a(\bar{y}_i)} \,,
\end{equation}
and since $\Delta f\geq0$, we conclude $a_R'(\bar{y}_i)<0$. If bulk matter is present in the interval then \eqref{eq:fmderfromstresstensorcons} implies that $f'(y)>0$ as $y\to \bar{y}_i^+$, thus contradicting the assumption that $f(\bar{y}_i^+)$ is a maximum. This option is therefore ruled out. A similar argument rules out the option that the maximum is reached as $y\to \bar{y}_{i+1}^-$ (for $i=0,\ldots,n-1$).
The last option to consider is the case where no bulk matter fields are turned on in the interval $[\bar{y}_i,\bar{y}_{i+1}]$. In this case, for $\bar{y}_i<y<\bar{y}_{i+1}$ we have $f(y)=f_\text{max}=\text{const}$. The same argument applied for $y\to \bar{y}_i^+$ and $y\to \bar{y}_{i+1}^-$ still means that $a_R'(\bar{y}_i)<0$ and $a_L'(\bar{y}_{i+1})>0$ respectively, which in turn implies there must be some $y_\ast \in (\bar{y}_i,\bar{y}_{i+1})$ such that $a'(y_\ast)=0$ and $f(y_\ast)=f_\text{max}$. We then choose $y_\ast=0$, by shift symmetry, and the rest of the proof follows.

\section{Bounds on the tension in the thin-brane model in general dimensions}
\label{app:thin-brane-tension-bounds}

The tension bounds for the thin-brane model in $d=2$ were discussed in, 
\eg \cite{Bachas:2020yxv,Bachas:2021fqo}. Here we repeat the argument of
Ref.~\cite{Bachas:2021fqo} in the $y$ coordinates of the main text and in general dimensions. A related one-sided
condition appears in the AdS/BCFT literature, see \eg \cite{Fujita:2011fp}.

Without loss of generality, let $l_L\leq l_R$. We denote the common value of
the warp factor \eqref{eq:thinay} at the brane, or equivalently the radius of the induced
AdS$_d$ geometry, by
\begin{equation}
    a_B\equiv a(0)
    =l_L\cosh\left(\frac{y_L^*}{l_L}\right)
    =l_R\cosh\left(\frac{y_R^*}{l_R}\right).
    \label{eq:thin-brane-radius}
\end{equation}
In particular, $a_B\geq l_R$. Using this relation, the second matching
condition in \eqref{eq:thin-match123} becomes
\begin{equation}
    \frac{\Sigma}{d-1}
    =\operatorname{sign}(y_L^*)
    \sqrt{\frac{1}{l_L^2}-\frac{1}{a_B^2}}
    +\operatorname{sign}(y_R^*)
    \sqrt{\frac{1}{l_R^2}-\frac{1}{a_B^2}}.
\end{equation}
Since the first square root is at least as large as the second one, positivity
of the tension requires $y_L^*>0$, while $y_R^*$ may have either sign. Hence
\begin{equation}
    \frac{\Sigma}{d-1}
    =\sqrt{\frac{1}{l_L^2}-\frac{1}{a_B^2}}
    \pm
    \sqrt{\frac{1}{l_R^2}-\frac{1}{a_B^2}},
    \label{eq:thin-brane-tension-branches}
\end{equation}
where the two signs correspond to $y_R^*>0$ and $y_R^*<0$, respectively.
The branches meet at $a_B=l_R$. As $a_B$ increases from $l_R$ to infinity,
the minus branch decreases from $\sqrt{l_L^{-2}-l_R^{-2}}$ to $l_L^{-1}-l_R^{-1}$, while the plus branch
increases from $\sqrt{l_L^{-2}-l_R^{-2}}$ to $l_L^{-1}+l_R^{-1}$. We therefore obtain by joining the ranges
\begin{equation}
    \left|\frac{1}{l_L}-\frac{1}{l_R}\right|
    \leq \frac{\Sigma}{d-1}
    \leq \frac{1}{l_L}+\frac{1}{l_R},
    \label{eq:thin-brane-tension-bounds}
\end{equation}
and a very similar analysis holds for $l_R\leq l_L$, leading to the same bound. 
This is precisely the bound quoted in the main text in equation \eqref{boundN}. 
For \(l_L\neq l_R\), the inequalities are strict for a finite-radius
\(\mathrm{AdS}_d\) brane and are saturated only in the flat-brane limit
\(a_B\to\infty\). When \(l_L=l_R\), the lower bound \(\Sigma=0\) corresponds to
the degenerate homogeneous-AdS configuration. 
The signs of $y_L^*$ and $y_R^*$ distinguish the upper and lower bounds. Both $y_L^*$ and $y_R^*$ tend to $+\infty$ at the upper
bound. At the lower bound, for $l_L<l_R$, 
$y_L^*\to+\infty$ and $y_R^*\to-\infty$ and the other way around for $l_R<l_L$.

The allowed tension range \eqref{eq:thin-brane-tension-bounds} also has a simple physical interpretation. Tensions
in its interior give a finite-radius \(\mathrm{AdS}_d\) geometry on the
brane, as in the Karch--Randall construction \cite{Karch:2000ct}. For
\(l_L\neq l_R\), saturation of either bound sends \(a_B\to\infty\), so
that the brane becomes flat. 
Increasing the tension above the upper bound gives a
\(\mathrm{dS}_d\), rather than \(\mathrm{AdS}_d\), geometry on the brane. 
The lower bound has a different physical
interpretation. Below it, the larger-radius AdS vacuum is semiclassically unstable to the nucleation of bubbles of the smaller-radius vacuum
\cite{Coleman:1980aw,Barbon:2010gn,Bachas:2021fqo}.

For the AdS$_3$ thin-wall model, reference \cite{Czech:2016nxc} advocated the stronger condition
$\Sigma\geq\sqrt{l_L^{-2}-l_R^{-2}}$ for $l_L<l_R$, which
would remove the branch with  $y_R^*<0$. As emphasized in
reference \cite{Bachas:2021fqo}, however, the radius variation used in that
argument changes the action by an amount proportional to the infinite
AdS$_2$ volume and is therefore not a normalizable wall fluctuation. We therefore do not impose this proposed additional restriction here. The authors of  \cite{Czech:2016nxc} do not prove an analogous restriction in general dimensions, although it is natural to expect it will take the form $\Sigma/(d-1)\geq\sqrt{l_L^{-2}-l_R^{-2}}$ for $l_L<l_R$.

\section{Thin-brane transmission from the Israel conditions}
\label{app:thin-brane-israel}

The thin-brane warp factor \eqref{eq:thinay} has a discontinuous derivative
$a'(y)$ at the brane location $y=0$, and the background Einstein equations
therefore contain a localized delta-function contribution. More explicitly,
the relevant background equation can be written as
\begin{equation}
    \frac{a''(y)}{a(y)}
    =
    \frac{\Theta(-y)}{l_L^2}
    +\frac{\Theta(y)}{l_R^2}
    -\frac{\Sigma}{d-1}\delta(y).
    \label{eq:thin-background-delta}
\end{equation}
A pillbox integral across $y=0$ gives the background Israel condition
\begin{equation}
    (d-1)\frac{a_L'(0)-a_R'(0)}{a_B}
    =
    \Sigma.
    \label{eq:thin-background-israelN}
\end{equation}
Together with continuity of $a(y)$, this reproduces the matching conditions
\eqref{eq:thin-match123}. The presence of a discontinuity makes it natural to ask whether
the derivation of the transmission coefficient
\eqref{eq:transmission_JanusN} in the main text remains valid in this case.
Here, we rederive the result by solving the relevant equations for the metric
perturbation $\beta(y)$, cf.\ \eqref{eq:bmodepert}, separately in the two
AdS$_{d+1}$ regions on either side of the brane and gluing the solutions using
the linearized Israel conditions \eqref{Israel1}. We show that this reproduces
the thin-brane transmission coefficient \eqref{eq:thin_trans}. This should be viewed as a consistency check rather
than as resolving any concrete issue with the original derivation. Indeed,
the thin-brane model is distributional by construction, since the matter
localized on the brane is itself described by a delta function.

Consider the perturbation \eqref{eq:bmodepert}, and denote its restrictions
to the two sides of the brane by $\beta_L(y)$ and $\beta_R(y)$. Away from
the brane, the bulk equation \eqref{Eq:bN} holds independently in each
region and can be written as
\begin{equation}
    \left(a^d\beta_S'\right)'=0,
    \qquad S=L,R.
    \label{eq:thin-beta-bulk}
\end{equation}
It remains to determine the matching conditions at the location of the brane. We use a single
coordinate $y$, increasing from the left region to the right.

No perturbation of the brane embedding is required when only the
$\beta(y)$ metric perturbation \eqref{eq:bmodepert} is turned on. This follows from time-reversal
symmetry, since the $\beta$ perturbation, which is proportional to
$dt\,d\zeta$, is odd under $t\rightarrow -t$, whereas a stationary
displacement of the brane 
$y_{\mathrm{brane}}=\epsilon\pi(\zeta)$ is invariant under
$t\rightarrow -t$ and is therefore even. Since the unperturbed background, including the constant-tension brane,
preserves time-reversal symmetry, the $\beta(y)$ perturbation decouples
from brane bending at linear order in $\epsilon$. As no separate source for the brane
displacement is introduced, we expect it to remain at $y=0$. 

With the brane fixed, the only nonvanishing linear corrections to the
induced metric and extrinsic curvature are their $t\zeta$ components.
All other components of the matching conditions reduce to the background
conditions and are therefore already satisfied. It is consequently
sufficient to consider the $t\zeta$ component of the induced-metric
continuity condition and of the linearized Israel condition.

Denoting the common value of the warp factor at the brane by
$a_B\equiv a_S(0)$, the $t\zeta$ component of the induced metric
\eqref{eq:bmodepert}, evaluated on either side of the brane, is
\begin{equation}
    \gamma_{t\zeta}^{\,S}
    =
    \epsilon a_B^2\beta_S(0)\zeta^{d-2},
    \qquad S=L,R.
\end{equation}
Continuity of the induced metric therefore requires
\begin{equation}
    \beta_L(0)=\beta_R(0)\equiv\beta_B.
    \label{eq:thin-beta-continuity}
\end{equation}
Since the perturbation \eqref{eq:bmodepert} has no $y$ components, its
contribution to the $t\zeta$ component of the extrinsic curvature is
\begin{equation}
        \delta K_{t\zeta}^{\,S}
        =
        -\frac{\epsilon}{2}
        \zeta^{d-2} \left.
        \partial_y\left(a_S^2\beta_S\right)
        \right|_{y=0}
        =
        -\epsilon \zeta^{d-2} a_B^2
        \left[
            \frac{a_S'(0)}{a_B}\beta_S(0)
            +\frac{1}{2}\beta_S'(0)
        \right].
    \label{eq:thin-delta-K}
\end{equation}
Here, $a_S'(0)$ denotes the one-sided derivative of the warp factor on
side $S$. As in equations \eqref{Eq:action}-\eqref{Israel1}, we use the same normal on both sides,
pointing towards decreasing $y$.

The Israel condition \eqref{Israel1} can easily be brought to the form
\begin{equation}
    [K_{\alpha\beta}]\equiv K_{\alpha\beta}^{\,L}-K_{\alpha\beta}^{\,R}
    =
    -\frac{\Sigma}{d-1}\gamma_{\alpha\beta}.
\end{equation}
Its linearized $t\zeta$ component is
\begin{equation}
\begin{split}
    0
    ={}&
    K_{t\zeta}^{\,L}
    -K_{t\zeta}^{\,R}
    +\frac{\Sigma}{d-1}\gamma_{t\zeta}
    \\={}&
    -\epsilon a_B^2\zeta^{d-2}
    \left[
        \left(
            \frac{a_L'(0)-a_R'(0)}{a_B}
            -\frac{\Sigma}{d-1}
        \right)\beta_B
        +\frac{1}{2}
        \left(\beta_L'(0)-\beta_R'(0)\right)
    \right]
   \\
    ={}&
    -\frac{\epsilon}{2}a_B^2\zeta^{d-2}
    \left(\beta_L'(0)-\beta_R'(0)\right),
\end{split}
\end{equation}
where the background Israel condition
\eqref{eq:thin-background-israelN} was used in the last equality. 
Combining this result with
\eqref{eq:thin-beta-continuity}, the matching conditions for $\beta$ are
\begin{equation}
    \beta_L(0)=\beta_R(0),
    \qquad
    \beta_L'(0)=\beta_R'(0).
    \label{eq:thin-beta-matching}
\end{equation}

We now solve the bulk equation \eqref{eq:thin-beta-bulk}, first
separately in the two AdS$_{d+1}$ regions and then impose the matching
conditions \eqref{eq:thin-beta-matching} at the brane. 
The Fefferman--Graham expansion identifies the constant asymptotic value of $\beta$ with the boundary metric source. 
To introduce a source $\beta_0^L$ on the left and no source on the right,
we therefore impose, as in \eqref{eq:asymsources}, 
\begin{equation}\label{eq:bcappendix}
    \lim_{y\to-\infty}\beta_L(y)=\beta_0^L,
    \qquad
    \lim_{y\to+\infty}\beta_R(y)=0.
\end{equation}
Integrating \eqref{eq:thin-beta-bulk} gives $a^d\beta_S'=-\tilde B_S$ for each side $S=L,R$, where $\tilde B_S$ are constants. Since $a(y)$ is continuous at
the brane, continuity of $\beta'$ gives
$\tilde B_L=\tilde B_R\equiv\tilde B$. Using the boundary conditions \eqref{eq:bcappendix} to fix
the additive integration constants, the two solutions can be written as
\begin{equation}
\begin{split}\label{eq:splitbeta1}
    \beta_L(y)
    &=
    \beta_0^L-\tilde B
    \int_{-\infty}^{y}a(y')^{-d}\,dy',
    \\
    \beta_R(y)
    &=
    \tilde B
    \int_{y}^{\infty}a(y')^{-d}\,dy'.
\end{split}
\end{equation}
Continuity of $\beta$ at $y=0$ then
requires
\begin{equation}
    \tilde B
    =
    \frac{\beta_0^L}
    {\displaystyle\int_{-\infty}^{\infty}a(y)^{-d}\,dy}.
\end{equation}
The matched solution can therefore be written in precisely the same
form as in the main text 
\eqref{eq:betaBsol},
\begin{equation}
    \beta(y)
    =
    \tilde B\int_y^\infty a(y')^{-d}\,dy',
    \label{eq:thin-beta-solution}
\end{equation}
where the integral is understood using the corresponding branch of
the piecewise warp factor \eqref{eq:thinay} on each side of the brane.

To relate this result to the asymptotic conventions used in the main
text, we now introduce separate shifted coordinates in the two patches,
\begin{equation}
    \hat y_L=y+y_L^*,
    \qquad
    \hat y_R=y-y_R^*.
    \label{eq:thin-asymptotic-coordinates}
\end{equation}
Their ranges are $\hat y_L\in(-\infty,y_L^*]$ and
$\hat y_R\in[-y_R^*,\infty)$; thus, they do not define a single
global coordinate system. Writing \eqref{eq:splitbeta1} in these
variables gives
\begin{align}
    \beta_L(\hat y_L)
    &=
    \beta_0^L-\tilde B
    \int_{-\infty}^{\hat y_L}
    \left[
        l_L\cosh\left(\frac{u}{l_L}\right)
    \right]^{-d}du,
    \nonumber\\
    \beta_R(\hat y_R)
    &=
    \tilde B
    \int_{\hat y_R}^{\infty}
    \left[
        l_R\cosh\left(\frac{v}{l_R}\right)
    \right]^{-d}dv.
    \label{eq:thin-beta-shifted}
\end{align}
These are precisely the patchwise forms to which the asymptotic expansion \eqref{Eq:expansion} and the subsequent Fefferman--Graham analysis apply, with $y$ replaced by $\hat y_L$ or $\hat y_R$. The argument following \eqref{Eq:expansion} therefore leads directly to the general transmission formula \eqref{eq:transmission_JanusN}. Splitting its integral at $y=0$ and substituting the two branches of \eqref{eq:thinay} reproduces the thin-brane result \eqref{eq:thin_trans}.

\section{The thin-brane calculation in angular coordinates}
\label{app:thin-brane-theta}

Angular coordinates adapted to the AdS$_d$ foliation are frequently used
in holographic descriptions of boundaries, interfaces, and branes, see \eg
\cite{Takayanagi:2011zk,Baig:2022cnb,Bachas:2022etu}. We therefore recast the
relevant thin-brane expressions directly in these coordinates.

For the two branches of the thin-brane warp factor
\eqref{eq:thinay}, we introduce separate angular coordinates by
\begin{equation}
    \tan\theta_L
    =
    \sinh\!\left(\frac{y+y_L^*}{l_L}\right),
    \qquad
    \tan\theta_R
    =
    \sinh\!\left(\frac{y_R^*-y}{l_R}\right).
    \label{eq:theta-y-map}
\end{equation}
The relative sign is chosen so that both angular coordinates have the same
local orientation, approaching $\theta_S=-\pi/2$ at the corresponding
asymptotic boundary and increasing toward the brane. Thus, on each side they
range over
\begin{equation}
    -\frac{\pi}{2}\leq\theta_S\leq\theta_S^B,
    \qquad S=L,R,
\end{equation}
where the brane is located at $\theta_S=\theta_S^B$. These definitions imply
\begin{equation}
    a(y)=\frac{l_S}{\cos\theta_S},
    \qquad
    dy
    =
    \begin{cases}
        \displaystyle
        \frac{l_L}{\cos\theta_L}\,d\theta_L
        =a(y)\,d\theta_L,
        & y<0,\\[7pt]
        \displaystyle
        -\frac{l_R}{\cos\theta_R}\,d\theta_R
        =-a(y)\,d\theta_R,
        & y>0.
    \end{cases}
    \label{eq:theta-y-differential}
\end{equation}
The background metric \eqref{Janus-metricN} therefore becomes
\begin{equation}
    ds_{(0),S}^2
    =
    \frac{l_S^2}{\cos^2\theta_S}
    \left[
        d\theta_S^2
        +\frac{-dt^2+d\zeta^2+dx_i^2}{\zeta^2}
    \right],
    \qquad S=L,R.
    \label{eq:theta-background}
\end{equation}
The background matching conditions \eqref{eq:thin-match123} take the form
\begin{equation}
    a_B
    \equiv
    \frac{l_L}{\cos\theta_L^B}
    =
    \frac{l_R}{\cos\theta_R^B},
    \qquad
    \frac{\Sigma}{d-1}
    =
    \frac{\tan\theta_L^B+\tan\theta_R^B}{a_B}.
    \label{eq:theta-background-matching}
\end{equation}
For comparison with the flat-sliced coordinates used in the asymptotic
analysis of the main text, cf.\ \eqref{eq:FGgeneric}, each angular patch can
also be mapped to Poincar\'e coordinates:
\begin{equation}
    \rho_L=\zeta\cos\theta_L,
    \qquad
    x=\zeta\sin\theta_L,
    \qquad\text{and}\qquad
    \rho_R=\zeta\cos\theta_R,
    \qquad
    x=-\zeta\sin\theta_R.
    \label{eq:theta-poincare-map}
\end{equation}
These are separate bulk extensions of the same signed boundary coordinate
$x$. In either patch,
\begin{equation}
    ds_{(0),S}^2
    =
    \frac{l_S^2}{\rho_S^2}
    \left[
        d\rho_S^2-dt^2+dx^2+dx_i^2
    \right].
    \label{eq:theta-poincare-metric}
\end{equation}
At $\rho_S=0$, the two patches cover the boundary half-spaces $x<0$ and
$x>0$, respectively.

The normal convention often used in the literature 
takes the normal covector to point toward increasing $\theta_S$ on both sides,
$n_S=+a_B\,d\theta_S$. In the  coordinate system used in the main text this maps to
$n_L=+dy$ and $n_R=-dy$.
Appendix~\ref{app:thin-brane-israel} instead uses the global normal $n=-dy$. Thus, the Israel matching condition \eqref{Israel1} is modified by $K_L\to-K_L$, with $K_R$ unchanged. The differences in \eqref{Israel1} therefore become minus the sums of the extrinsic curvature contributions.

The perturbation \eqref{eq:bmodepert} takes the form
\begin{equation}
    ds_S^2
    =
    \frac{l_S^2}{\cos^2\theta_S}
    \left[
        d\theta_S^2
        +\frac{-dt^2+d\zeta^2+dx_i^2}{\zeta^2}
        +2\epsilon\,b_S(\theta_S)\zeta^{d-2}dt\,d\zeta
    \right],
    \label{eq:theta-beta-perturbation}
\end{equation}
where $b_S(\theta_S)\equiv\beta_S(y(\theta_S))$.
Using \eqref{eq:theta-y-differential}, the bulk equation for the perturbation \eqref{Eq:bN} (equivalently \eqref{eq:thin-beta-bulk})
becomes, on either side,
\begin{equation}
    \partial_{\theta_S}
    \left(
        \sec^{d-1}\!\theta_S\,
        \partial_{\theta_S}b_S
    \right)
    =0.
    \label{eq:theta-beta-eom}
\end{equation}

The matching conditions derived in the preceding appendix become
\begin{equation}
    b_L(\theta_L^B)=b_R(\theta_R^B),
    \qquad
    b_L'(\theta_L^B)=-b_R'(\theta_R^B),
    \label{eq:theta-beta-matching}
\end{equation}
where 
primes on functions of $\theta_S$ denote angular
derivatives. 
The minus sign in the second condition follows from the opposite orientations
of the two angular coordinate systems in the two patches. Indeed, \eqref{eq:theta-y-differential} gives
$\partial_y\theta_L|_B=1/a_B$ and
$\partial_y\theta_R|_B=-1/a_B$. In the global $y$ coordinate, the same
condition is simply
$\partial_y\beta_L(0)=\partial_y\beta_R(0)$.

To solve \eqref{eq:theta-beta-eom}, define
\begin{equation}
    \mathcal B(\theta)
    \equiv
    \int_{-\pi/2}^{\theta}
    \cos^{d-1}\!\vartheta\,d\vartheta,
    \qquad
    \mathcal B'(\theta)=\cos^{d-1}\!\theta.
    \label{eq:theta-B-function}
\end{equation}
At the brane,
\begin{equation}
    \mathcal B'(\theta_S^B)
    =
    \left(\frac{l_S}{a_B}\right)^{d-1}.
    \label{eq:theta-B-prime-at-brane}
\end{equation}
We impose the same left-source boundary conditions as in
\eqref{eq:asymsources}, namely, the source is $\beta_0^L$ on the left and vanishes on
the right. Together with \eqref{eq:theta-B-prime-at-brane}, the derivative
matching condition allows the two solutions to be written in terms of a
common constant $\tilde B$ as
\begin{equation}
    b_L(\theta_L)
    =
    \beta_0^L
    -\frac{\tilde B}{l_L^{d-1}}\mathcal B(\theta_L),
    \qquad
    b_R(\theta_R)
    =
    \frac{\tilde B}{l_R^{d-1}}\mathcal B(\theta_R).
    \label{eq:theta-left-source-solution}
\end{equation}
The remaining continuity condition fixes $\tilde B$ to be
\begin{equation}
    \tilde B
    =
    \frac{\beta_0^L}{
        l_L^{-(d-1)}\mathcal B(\theta_L^B)
        +l_R^{-(d-1)}\mathcal B(\theta_R^B)
    }.
    \label{eq:theta-Btilde}
\end{equation}
This is exactly the solution
\eqref{eq:betaBsol} up to the angular change of coordinates in each patch.

The stress-tensor on the right boundary can also be read off directly in these
coordinates. Let
$u_S\equiv\theta_S+\pi/2$. Near either asymptotic boundary, $\mathcal B(\theta_S)=\frac{u_S^d}{d}+\mathcal O(u_S^{d+2})$. Near the right boundary we therefore have $b_R
    =
    \frac{\tilde B}{d\,l_R^{d-1}}u_R^d
    +\mathcal O(u_R^{d+2})$. 
The right-hand solution therefore contains only the normalizable mode. Near
the right boundary, \eqref{eq:theta-poincare-map} gives
$u_R=\rho_R/x+\cdots$ and $\zeta=x+\cdots$. After the standard subleading
adjustment to Fefferman--Graham gauge, its leading contribution is
\begin{equation}
    ds_R^2
    =
    ds_{(0),R}^2
    +2\epsilon\,\rho_R^{d-2}
    \frac{\tilde B}{d\,l_R^{d-3}}dt\,dx
    +\cdots.
    \label{eq:theta-FG-perturbation}
\end{equation}
Hence, the  relevant order-$d$ Fefferman--Graham
coefficient is
$g_{tx}^{(d)}
    =
    \frac{\epsilon\tilde B}{d\,l_R^{d-1}}$.
Using
$\langle T_{ij}\rangle
=d\,l_R^{d-1}g_{ij}^{(d)}/(16\pi G_N)$
\cite{deHaro:2000vlm} this gives
\begin{equation}
    \left\langle T_{tx}^{R}\right\rangle_I
    =
    \frac{\epsilon\tilde B}{16\pi G_N}.
    \label{eq:theta-transmitted-stress}
\end{equation}
By continuity of the energy flux,
$\langle T_{tx}^{L}\rangle_I=\langle T_{tx}^{R}\rangle_I$.

Finally, the homogeneous left reference with the same source is obtained
from the no-brane limit of \eqref{eq:theta-Btilde} and
\eqref{eq:theta-transmitted-stress}. In this limit,
$l_R=l_L$ and $\theta_R^B=\theta_L^B=0$. We therefore obtain the
transmission coefficient
\begin{equation}
    \mathcal T_L
    =
    \frac{\langle T_{tx}^R\rangle_I}
         {\langle T_{tx}^L\rangle_0}
    =
    \frac{2l_L^{-(d-1)}\mathcal B(0)}
    {l_L^{-(d-1)}\mathcal B(\theta_L^B)
     +l_R^{-(d-1)}\mathcal B(\theta_R^B)}.
    \label{eq:theta-y-equivalence}
\end{equation}
Changing variables in the integrals defining $\mathcal B$ according to
$\tan\vartheta=\sinh\tilde y$ reproduces precisely the thin-brane
result \eqref{eq:thin_trans}.

\section{Explicit thin-brane results in different dimensions}
\label{app:thin-brane-explicit}

The general thin-brane result \eqref{eq:thin_trans}, or \eqref{eq:theta-y-equivalence} in angular coordinates, can be evaluated in terms of 
elementary functions for each integer $d$. To make the dependence on the
dimension explicit, we write the function defined in
\eqref{eq:theta-B-function} as
\begin{equation}
    \mathcal B_d(\theta)
    \equiv
    \int_{-\pi/2}^{\theta}
    \cos^{d-1}\!\vartheta\,d\vartheta.
    \label{eq:explicit-B-definition}
\end{equation}
Under the change of variables $\tan\theta=\sinh\tilde y$, the function
$\tau$ appearing in \eqref{eq:thin_trans} is simply
\begin{equation}
    \tau(\tilde y)
    =
    \frac{\mathcal B_d(\theta)}{\mathcal B_d(0)}.
    \label{eq:explicit-tau-B}
\end{equation}

The functions \eqref{eq:explicit-B-definition} obey the recurrence relation
\begin{equation}
    \mathcal B_{d+2}(\theta)
    =
    \frac{\sin\theta\cos^d\!\theta}{d+1}
    +\frac{d}{d+1}\mathcal B_d(\theta).
    \label{eq:explicit-B-recurrence}
\end{equation}
For $d=2,\ldots,6$, they are
\begin{subequations}
\label{eq:explicit-B-low-dimensions}
\begin{align}
    d=2:\qquad
    &\mathcal B_2(\theta)
    =1+\sin\theta,
    \qquad
    2\mathcal B_2(0)=2,
    \label{eq:explicit-B-d2}
    \\
    d=3:\qquad
    &\mathcal B_3(\theta)
    =\frac{\pi+2\theta+\sin(2\theta)}{4},
    \qquad
    2\mathcal B_3(0)=\frac{\pi}{2},
    \label{eq:explicit-B-d3}
    \\
    d=4:\qquad
    &\mathcal B_4(\theta)
    =\frac{2+3\sin\theta-\sin^3\theta}{3},
    \qquad
    2\mathcal B_4(0)=\frac{4}{3},
    \label{eq:explicit-B-d4}
    \\
    d=5:\qquad
    &\mathcal B_5(\theta)
    =\frac{6\pi+12\theta+8\sin(2\theta)+\sin(4\theta)}{32},
    \qquad
    2\mathcal B_5(0)=\frac{3\pi}{8},
    \label{eq:explicit-B-d5}
    \\
    d=6:\qquad
    &\mathcal B_6(\theta)
    =\frac{8+15\sin\theta-10\sin^3\theta+3\sin^5\theta}{15},
    \qquad
    2\mathcal B_6(0)=\frac{16}{15}.
    \label{eq:explicit-B-d6}
\end{align}
\end{subequations}
More generally,
\begin{equation}
    2\mathcal B_d(0)
    =
    \frac{\sqrt{\pi}\,\Gamma(d/2)}{\Gamma((d+1)/2)}.
    \label{eq:explicit-B-zero}
\end{equation}

We now introduce the AdS radius ratio and the dimensionless brane tension 
\begin{equation}
    L\equiv\frac{l_L}{l_R},
    \qquad
    T\equiv\frac{\Sigma \, \sqrt{l_L l_R}}{d-1}.
    \label{eq:explicit-LT-definition}
\end{equation}
In terms of the brane angles, \eqref{eq:thin_trans} becomes
\begin{equation}
    \mathcal T_L
    =
    \frac{2\mathcal B_d(0)}{
        \mathcal B_d(\theta_L^B)
        +L^{d-1}\mathcal B_d(\theta_R^B)
    }.
    \label{eq:explicit-angular-transmission}
\end{equation}
The background matching conditions \eqref{eq:theta-background-matching}
give 
\begin{equation}
    \sin\theta_L^B
    =
    \frac{\sqrt{L}}{2T}\left(T^2+1/L-L\right),\qquad 
    \sin\theta_R^B
    =
    \frac{1}{2T\sqrt{L}}\left(T^2+L-1/L\right),
    \label{eq:explicit-angle-LT-right}
\end{equation}
where $\theta_L^B,\theta_R^B\in[-\pi/2,\pi/2]$. Thus, the even-$d$ results are algebraic in $L$
and $T$, while the odd-$d$ results also contain inverse trigonometric
functions.

For example, for $d=2$, corresponding to three bulk dimensions,
\eqref{eq:explicit-angular-transmission} reduces to
\begin{equation}
    \mathcal T_L^{(d=2)}
    = \frac{2}{1+L+\sin(\theta_L^B)+L\sin(\theta_R^B)} = 
    \frac{2}{1+L+T\sqrt{L}},
    \label{eq:explicit-transmission-d2}
\end{equation}
in agreement with the known two-dimensional result \cite{Bachas:2020yxv}. For $d=3$,
corresponding to four bulk dimensions, define 
\begin{equation}
    \Xi
    \equiv
    L\sqrt{
        \left[\left(\frac{1}{\sqrt L}+\sqrt{L}\right)^2-T^2\right]
        \left[T^2-\left(\frac{1}{\sqrt{L}}-\sqrt{L}\right)^2\right]
    }.
    \label{eq:explicit-S-definition}
\end{equation}
Then 
\begin{equation}
    \mathcal T_L^{(d=3)}
    =
    \frac{2\pi}{
        \pi(1+L^2)+\Xi
        +2\sin^{-1}\!\left(\frac{\sqrt{L}}{2T}\left(T^2+\frac{1}{L}-L\right)\right)
        +2L^2\sin^{-1}\!\left(\frac{1}{2T\sqrt{L}}\left(T^2+L-\frac{1}{L}\right)\right)
    }.
    \label{eq:explicit-transmission-d3}
\end{equation}

Finally, in terms of the dimensionless brane tension and AdS radius ratio, the tension range \eqref{boundN} becomes
\begin{equation}
\left|\frac{1}{\sqrt{L}}-\sqrt{L}\right|
\leq T \leq
\frac{1}{\sqrt{L}}+\sqrt{L}\,,
    \label{eq:explicit-tension-range}
\end{equation}
and the corresponding transmission coefficient obeys \eqref{eq:thin_tension_bounds}:
\begin{equation}
    \frac{1}{1+L^{d-1}}
    \leq
    \mathcal T_L
    \leq
    \min\!\left(1,L^{1-d}\right).
    \label{eq:explicit-transmission-range}
\end{equation}

As an interesting side remark, we can take the tensionless
empty-$\mathrm{AdS}$ limit by approaching $(L,T)=(1,0)$ along the
equal-radius family $L=1$, $T\to0$. Along this path, both brane
angles \eqref{eq:explicit-angle-LT-right} vanish, and the transmission coefficient
\eqref{eq:explicit-angular-transmission} approaches 1, as expected.
There are, however, other admissible paths approaching the same point
while satisfying \eqref{eq:explicit-tension-range}, which may yield
nonzero, path-dependent limiting angles. Nevertheless, the condition \eqref{eq:explicit-tension-range}
forces the two angles to become opposite in the limit, namely,
$\theta_R^B=-\theta_L^B$. Since
$L^{d-1}\to1$, the corresponding angular integrals in
\eqref{eq:explicit-angular-transmission}  become complementary,
and their sum approaches the pure-$\mathrm{AdS}$ value. Consequently, the limiting transmission is path-independent,
with $\mathcal{T}_L^{(d)}\to1$ along every such path.

\begin{figure*}[t]
    \centering
    \includegraphics[width=0.435\linewidth,trim={0cm 1.42cm 6cm 1cm},clip]{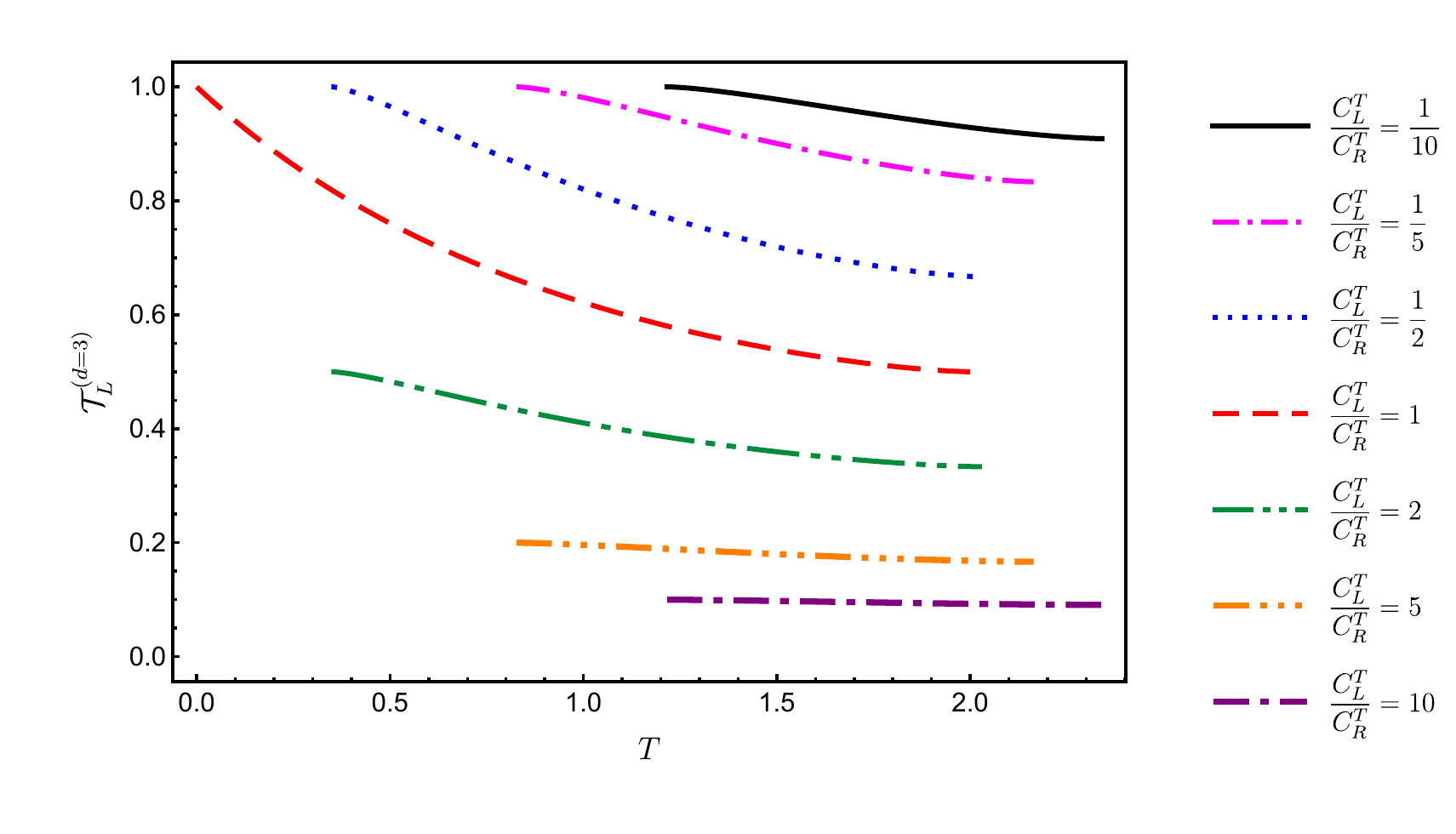}
     \hfill \includegraphics[width=0.55\linewidth,trim={0cm 1.1cm 0cm 0.8cm},clip]{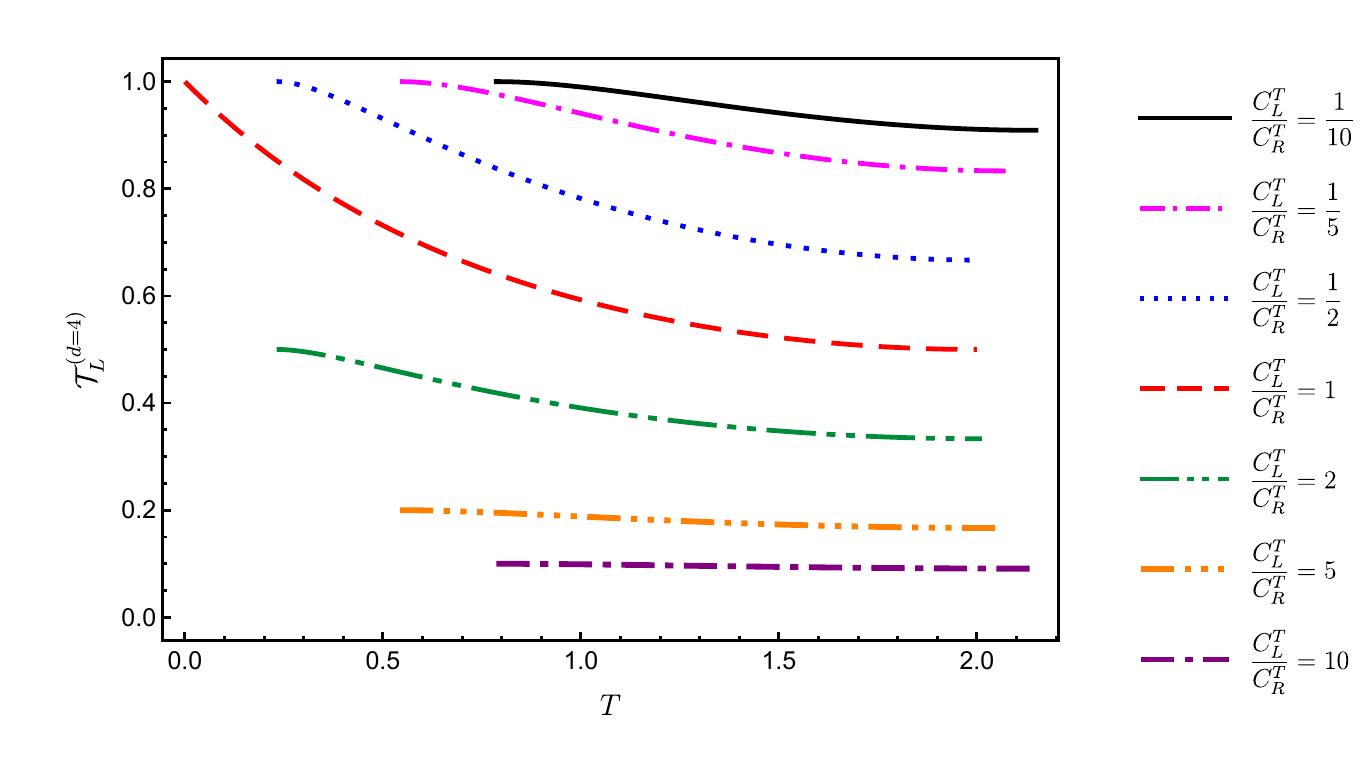}
    \caption{The thin-brane transmission coefficient $\mathcal T_L$ as a function of the dimensionless tension parameter $T$, defined in Eq.~\eqref{eq:explicit-LT-definition}, for several values of $C_L^T/C_R^T$. Left: $d=3$. Right: $d=4$. Each curve is shown over the admissible tension interval \eqref{eq:explicit-tension-range}. The different curves correspond to  different values of the ratio $C_L^T/C_R^T$ as specified in the legend.}\label{fig:assymappendix}
\end{figure*}

Figure~\ref{fig:assymappendix} complements the $d=3$ contour plot in figure \ref{fig:thin-brane} in the main text by showing the transmission coefficient as a function of the dimensionless tension parameter for several fixed values of the ratio $C_L^T/C_R^T=L^{d-1}$ in dimensions $d=3,4$. For each ratio, the transmission decreases monotonically across the admissible tension interval, from $\min(1,C_R^T/C_L^T)$ at the lower tension bound to $C_R^T/(C_L^T+C_R^T)$ at the upper tension bound, cf. equation \eqref{eq:thin_tension_bounds}. Thus, full transmission at the lower endpoint is possible when $C_L^T\leq C_R^T$, whereas for $C_L^T>C_R^T$ the maximal transmission is $C_R^T/C_L^T<1$. At fixed $C_L^T/C_R^T$, the endpoint values are independent of the dimension, while the admissible tension interval and the precise shape of the curves depend on $d$.

\section{Transmission coefficient in $d=2$ Janus geometry}

In this appendix, we  present two different derivations of the transmission coefficient for the specific case of a smooth Janus geometry in $d=2$ dimensions, with a single background scalar and a flat potential of the form $V(\phi)=-1/l^2$.  
We start by deriving and solving  the equations of motion obeyed by the transverse traceless perturbation of the background metric, and demonstrating that it is consistent to treat it without turning on the other scalar and vector sectors. 
Then, in Sec.~\ref{app:Januspart1}, we implement the method described in
Sec.~\ref{sec:holosetup} for finding the transmission coefficient in this
specific nontrivial example at zero frequency, demonstrating in greater
detail the derivation outlined in the main text. In particular, we allow
for a general static transverse-traceless metric source rather than
restricting to the off-diagonal perturbation \eqref{eq:bmodepert}.
Finally, in Sec.~\ref{app:Januspart2}, we reproduce and clarify the
calculation of Ref.~\cite{Bachas:2022etu}, keeping the frequency $\omega$
of the propagating wave generic. Two Fefferman--Graham patches, one on
each side of the interface, are constructed so that the boundary sources
vanish separately in each patch, and the corresponding coordinate
transformations are then matched at the interface.

The model we consider corresponds to the action \eqref{Janus action} with $d=2$. The unperturbed metric takes the form \eqref{Janus-metricN}
\begin{equation}
ds_{(0)}^2 = dy^2 + a^2(y) \left(\frac{d\zeta^2-dt^2}{\zeta^2}\right),
\end{equation}
where the warp factor $a(y)$ and the scalar field obey the equations of motion \eqref{eq:Janus}: 
\begin{equation}\label{eq:janus G}
\begin{split}
\left[
\left(\frac{a'}{a}\right)^2 + \frac{1}{a^2}
\right]
& =
\frac{1}{2}(\phi')^2 - V(\phi),
\qquad
 \phi'' + 2 \phi' \frac{a'}{a} - \frac{dV}{d\phi}  = 0 \, .
\end{split}
\end{equation}

Before imposing the flat-potential restriction, we keep the two
asymptotic AdS$_3$ radii $l_L$ and $l_R$ distinct. The transmission
coefficient can be written in the form used in
Ref.~\cite{Bachas:2022etu} as (note that this is only true in $d=2$) 
\begin{equation}
\mathcal{T}^{(d=2)}_L=
\frac{2/l_L}{
 \frac{1}{l_L}+\frac{1}{l_R}
+\int_{-\infty}^{\infty} \phi'(y)^2 dy}
=
\frac{
\int_{-\infty}^{\infty}\frac{1}{a_L(y)^2}dy }
{
\int_{-\infty}^{\infty}\frac{1}{a(y)^2}dy }.
\label{eq:JanusTLd2}
\end{equation}
The second equality follows from the background equations of motion,
which imply
\begin{equation}
\frac{1}{a(y)^{2}}
=
\partial_y \left(\frac{a'(y)}{a(y)}\right)+\phi'(y)^2,
\end{equation}
together with
$a'/a\to-1/l_L$ as $y\to-\infty$,
$a'/a\to1/l_R$ as $y\to+\infty$, and
$\int_{-\infty}^{\infty}dy\,a_L(y)^{-2}=2/l_L$.
The final expression is precisely the general result
\eqref{eq:transmission_JanusN} specialized to $d=2$. 

For the remainder of this appendix, we specialize to the flat
potential $V(\phi)=-1/l^2$, in which case the two asymptotic radii
coincide, $l_L=l_R=l$, and the pure-AdS reference warp factors are
identical on the two sides:  $a_L(y)=a_R(y)=l\cosh\left(\frac{y}{l}\right)$.
For this potential, the second equation in \eqref{eq:janus G}
implies that $\phi'(y)a(y)^2$ is constant. Substituting this
relation into the first equation in \eqref{eq:janus G} gives
\begin{align}\label{Janus background a}
        a(y)&= \frac{l}{\sqrt{2}}\bigg[ 1+(1-b)\cosh\left( \frac{2y}{l} \right) \bigg]^{1/2}\,, \qquad 
        \phi(y)=\phi_0 + \frac{1}{\sqrt{2}} \log\left[ \frac{\sqrt{2-b}+\sqrt{b}\tanh(y/l)}{\sqrt{2-b}-\sqrt{b}\tanh(y/l)}   \right] \,,
\end{align}
where the Janus parameter $b \in \left[0,1\right]$. 
Here we follow the convention of \cite{Bachas:2022etu} for the Janus parameter $b$, corresponding to $b_{\text{there}}$ in footnote~\ref{foot:bbhatconventions}, rather than the $b$ used in Eq.~\eqref{phi_function} and the rest of the main text. 
It  is related to the jump in the dilaton source across the interface by: 
\begin{equation}
\Delta\phi  =\phi_{\infty}-\phi_{-\infty}= \sqrt{2}\log\left( \frac{1+\sqrt{b(2-b)}}{1-b}\right)\,.
\end{equation} 
When $b=0$, the bulk is pure AdS$_3$, while for $b=1$, the dilaton is linear and the  geometry becomes $\mathbb{R}\times \text{AdS}_2$. 
As first reported in \cite{Bachas:2022etu}, the transmission coefficient across the interface of the dual ICFT is:
\begin{equation}\label{eq:JanusTransmission2dappG}
\mathcal{T}_L=\mathcal{T}_R=\frac{1}{2}\sqrt{b(2-b)}\left[\arctanh \left(\sqrt{\frac{b}{2-b}}\right) \right]^{-1}\,.
\end{equation}
Indeed, substituting the warp factor in \eqref{Janus background a}, together with
$a_L(y)=l\cosh(y/l)$, into \eqref{eq:JanusTLd2} gives this result.

In what follows, we rederive the above result for the transmission coefficient in two different ways. As a common starting point, we consider the linearized equations of motion for general, small (order $\epsilon$) perturbations of the AdS$_2$ slice metric and scalar field. We then show that the scalar perturbation, together with the trace and divergence of the metric perturbation, can consistently be set to zero, and solve the remaining equations for the transverse-traceless metric perturbation, used in both derivations.

We denote the unit-radius AdS$_2$ metric by
$\gamma_{\alpha\beta}dx^\alpha dx^\beta
\equiv (d\zeta^2-dt^2)/\zeta^2$, with
$x^\alpha=(\zeta,t)$. Working in radial gauge,
$\delta g_{yy}=\delta g_{y\alpha}=0$, we parametrize the perturbations
of the slice metric and scalar field by $h_{\alpha\beta}$ and
$\tilde{\phi}_1$, respectively:
\begin{subequations}\label{janus perturbations}
    \begin{align}\label{janus perturbations metric}
        ds^2&=dy^2 + a(y)^2\left[ \gamma_{\alpha\beta} + \epsilon h_{\alpha\beta}(y,\zeta, t)\right]dx^\alpha dx^\beta\,,\\
        \Phi(y,\zeta, t)&=\phi(y)+ \epsilon \tilde{\phi_1}(y,\zeta, t)\,.
    \end{align}%
\end{subequations}
Because the background is invariant under time translations, Fourier
modes of different frequencies decouple in the linearized equations. We may therefore work with
a single Fourier mode of the form
\begin{equation}\label{eq:metricpertgenapp}
    h_{\alpha\beta}(y,\zeta,t)=\frac{e^{i\omega  t}}{l^2} \left(
\begin{array}{cc}
 H_\zeta(y,\zeta ) +\frac{l^2}{2\zeta^2}\kappa(y,\zeta )& 
 H_t(y,\zeta ) \\
 H_t(y,\zeta ) & 
 H_\zeta(y,\zeta ) - \frac{l^2}{2\zeta^2}\kappa(y,\zeta ) \\
\end{array}
\right)\,, \qquad 
\tilde{\phi_1}(y,\zeta, t)=  \, e^{i\omega t}\phi_1(y,\zeta).
\end{equation}

To organize the perturbations according to their tensorial character on
the AdS$_2$ slices, we use the scalar fluctuation $\phi_1$, the metric
trace $\kappa$, and the covariant divergence $V_\beta$ of
$h_{\alpha\beta}$:
\begin{subequations}\label{eq:SVT}
\begin{align}
\text{Scalar:}& \qquad
\left\{
\phi_1,\;
\gamma^{\alpha\beta}h_{\alpha\beta}
=
e^{i\omega t}\kappa(y,\zeta)
\right\}
\label{eq:SVTscalar}
\\[1ex]
\text{Vector:}& \qquad
\nabla^{(\gamma)}{}^\alpha h_{\alpha\beta}
\equiv
e^{i\omega t}
\begin{pmatrix}
V_\zeta(y,\zeta)\\
V_t(y,\zeta)
\end{pmatrix}
=
e^{i\omega t}
\begin{pmatrix}
\frac{\zeta^2}{l^2}\left(-i\omega H_t+\partial_\zeta H_\zeta\right)+\frac{1}{2}\partial_\zeta \kappa\\
\frac{\zeta^2}{l^2}\left(-i\omega H_\zeta+\partial_\zeta H_t\right)+\frac{i\omega}{2}\kappa
\end{pmatrix}
\label{eq:SVTvector}
\\
\text{Tensor:}& \qquad
h^{TT}_{\alpha\beta}: \quad \nabla^{(\gamma)\alpha }h^{TT}_{\alpha\beta}=0 \,, \quad \gamma^{\alpha\beta} h^{TT}_{\alpha \beta}=0\,.
\label{eq:SVTtensor}
\end{align}
\end{subequations}
Here and below, $\gamma_{\alpha\beta}$ is used to raise and lower indices
and to define all contractions and covariant derivatives. 
The quantity $V_\beta$ is determined by $h_{\alpha\beta}$ through
\eqref{eq:SVTvector} and is therefore not an independent field.
Imposing $V_\beta=0$ restricts the metric perturbation to the transverse
sector, while imposing $\kappa=0$ in addition makes it
transverse-traceless.

Substituting the parametrization \eqref{eq:metricpertgenapp} into the
Klein--Gordon and Einstein equations, expanding to linear order in
$\epsilon$, and using the background equations \eqref{eq:janus G}, we
obtain the following coupled system:\footnote{For a related decomposition of the linearized equations of
motion, see Eqs.~(4.19)--(4.23) of
Ref.~\cite{Papadimitriou:2004rz}. That analysis uses Euclidean
Fefferman-Graham coordinates and works at leading nontrivial order in
the Janus deformation. Here instead we use Lorentzian AdS$_2$ slicing
and the exact Janus background.}
\begin{equation}
E_\Phi:\quad
2\left[
\frac{\zeta^{2}\omega^{2}
+\zeta^2\partial_{\zeta}^{2}}{a(y)^2}
+\partial_{y}^{2}+
2\frac{a'(y)}{a(y)}\partial_{y}
\right]\phi_{1}
+
\phi'(y)\partial_{y}\kappa
=0,
\end{equation}
\begin{subequations}\label{Janus Einstein pert}
\begin{align}\label{Janus Einstein pert yy}
E_{yy} :\quad
&
\left(
\zeta^2\partial_\zeta^2
+\zeta^2\omega^2-1+a(y)a'(y)\,\partial_y\right)\kappa(y,\zeta)
-2a(y)^2\phi'(y)\,\partial_y\phi_1(y,\zeta)
\\ & \nonumber
\qquad \qquad\qquad \qquad\qquad \qquad\qquad+\zeta^2\left[
i\omega V_t(y,\zeta)
-\partial_\zeta V_\zeta(y,\zeta)
\right]
=0,
\\
\label{Janus Einstein pert yz}
E_{y\zeta } :\quad
&
-\phi'(y) \partial_\zeta \phi_1(y,\zeta)-\frac{1}{2} \partial_y \partial_\zeta \kappa(y,\zeta) +\frac{1}{2}\partial_y V_\zeta 
=0,
\\
\label{Janus Einstein pert yt}
E_{yt}:\quad
&
-i\omega \phi'(y) \phi_1(y,\zeta)-\frac{i\omega}{2}  \partial_y  \kappa(y,\zeta) +\frac{1}{2}\partial_y V_t 
=0,
\\[1ex]\label{Janus Einstein pert tt}
E_{\zeta\zeta}-E_{tt}:\quad
&
\left[\partial_y^2 + 2\frac{a'(y)}{a(y)}
\,\partial_y
\right]\kappa(y,\zeta)+
4\phi'(y)
\partial_y \phi_1(y,\zeta)=0,
\\[1ex]\label{Janus Einstein pert zz}
E_{\zeta\zeta}+E_{tt}:\quad
&
\left[
\partial_y^2
+2\frac{a'(y)}{a(y)}
\,\partial_y
\right]H_\zeta(y,\zeta)=0,
\\[1ex]\label{Janus Einstein pert zt}
E_{\zeta t}:\quad
&
\left[
\partial_y^2
+2\frac{a'(y)}{a(y)}
\,\partial_y
\right]H_t(y,\zeta)=0.
\end{align}
\end{subequations}
Because of the AdS$_2$ scaling symmetry, when $\omega\neq 0$, all equations of motion can be written in terms of $\tilde{\zeta}=\omega\zeta$, provided one also redefines $\tilde V_\alpha=V_\alpha/\omega$ and $\tilde H_\alpha=H_\alpha/\omega^2$, while leaving $\kappa$ and $\phi_1$ unchanged. 
In terms of these rescaled variables, the equations of motion contain
no explicit dependence on $\omega$. Consequently, solutions at
different nonzero frequencies are related by the above rescaling.
This explains the fact that the previous computation in $d=2$ \cite{Bachas:2022etu} took $\omega \neq 0$ and got a result that does not depend on $\omega$.\footnote{One can also obtain the $\omega=0$ solution from these $\omega\neq 0$ solutions using the same scaling argument. However, this requires care when taking the limit $\omega \rightarrow 0$.}

The equations of motion lead to two further observations that will be useful
in the calculations of sections \ref{app:Januspart1}
and~\ref{app:Januspart2}. First, at $\omega=0$,
\eqref{eq:SVTvector} reduces to
\begin{equation}
V_\zeta
=
\frac{\zeta^2}{l^2}\partial_\zeta H_\zeta
+\frac{1}{2}\partial_\zeta\kappa,
\qquad
V_t
=
\frac{\zeta^2}{l^2}\partial_\zeta H_t .
\end{equation}
With these substitutions, together with $\omega=0$, the equations of motion split into two independent sets: one involving
$\{\phi_1,\kappa,H_\zeta\}$ and another involving only $H_t$, through
$E_{yt}$ and $E_{\zeta t}$. Exciting either set does not source the
other. This is the precise sense in which $H_t$, which parametrizes
$h_{t\zeta}$, decouples at zero frequency. 
It provides the explicit
$d=2$ realization of the symmetry-based decoupling argument given in
Sec.~\ref{sec:holosetup} for the perturbation
\eqref{eq:bmodepert}, which is further demonstrated in
Sec.~\ref{app:Januspart1}.

For the nonzero-frequency calculation of
Sec.~\ref{app:Januspart2}, we instead choose to restrict to the
transverse-traceless sector by imposing
\begin{subequations}\label{eq:TTtruncation}
\begin{align}
\phi_1\equiv\kappa&\equiv0,
\\
V_\zeta\equiv V_t&\equiv0.
\end{align}
\end{subequations}
This restriction is consistent with the equations of motion:
$E_\Phi$, $E_{yy}$, $E_{y\zeta}$, $E_{yt}$, and
$E_{\zeta\zeta}-E_{tt}$ are then identically satisfied, leaving
$E_{\zeta\zeta}+E_{tt}$ and $E_{\zeta t}$. 
It therefore selects a consistent subset of solutions, but does not
establish that the transverse-traceless response is unaffected by
arbitrary scalar or trace sources at nonzero frequency. 
A full decoupling, in the stronger sense that the physical scalar and
transverse-traceless modes obey independent linearized equations, was established for five-dimensional Einstein--dilaton backgrounds in
Ref.~\cite{ghodsi2025spectraholographicqftsconstant}, using an analogous
decomposition. 
Care is required in such an analysis because exceptional modes in the kernels of the relevant differential operators may prevent a unique decomposition into different sectors.\footnote{For example, on a $d$-dimensional constant-curvature
space with
$R_{\alpha\beta}= R\gamma_{\alpha\beta}/d$,  consider the
scalar-derived tensor
$X_{\alpha\beta}[q]
\equiv
\left(
\nabla_\alpha\nabla_\beta
-\frac{1}{d}\gamma_{\alpha\beta}\Box_\gamma
\right)q$.
It is traceless and satisfies
$\nabla^\alpha X_{\alpha\beta}[q]
=
\frac{d-1}{d}\nabla_\beta
\left(\Box_\gamma+R/(d-1)\right)q
$.
Consequently, when
$(\Box_\gamma+R/(d-1))q=0$, the tensor $X_{\alpha\beta}[q]$ is both
scalar-derived and transverse-traceless. In a decomposition containing
$h_{\alpha\beta}
=h_{\alpha\beta}^{TT}+X_{\alpha\beta}[\sigma]+\cdots$, the shifts
$\sigma\rightarrow\sigma+q$ and
$h_{\alpha\beta}^{TT}\rightarrow
h_{\alpha\beta}^{TT}-X_{\alpha\beta}[q]$
leave the full metric perturbation unchanged. Boundary or regularity
conditions are therefore required to fix this ambiguity.} 
It would be interesting to extend this analysis to the present $d=2$
setting. This case is additionally degenerate: at the level of local
component counting, the number of independent transverse-traceless
components is
$\frac{d(d+1)}{2}-d-1=\frac{(d+1)(d-2)}{2}$, which 
vanishes for $d=2$. The role of exceptional homogeneous modes and
boundary conditions is therefore especially important. Due to these extra complications, we choose here to leave this problem for future work and instead settle for simply imposing the above consistent restrictions in \eqref{eq:TTtruncation}.

Accordingly, in both calculations below we focus on
$h_{\alpha\beta}^{TT}$.
At $\omega=0$, this contains the mode $H_t$ from \eqref{eq:bmodepert} that we have shown to be decoupled from the others (as well as the additional mode $H_\zeta$\footnote{Even though for $\omega=0$, we could keep only $H_t$, we still keep $H_\zeta$ in this case for two different reasons. First, this allows us to smoothly take the $\omega\to 0$ limit. Second, it allows us to demonstrate certain subtleties about anomalous contributions to the stress tensor, as will be shown in \ref{app:Januspart1}.}). 
As discussed above, at general $\omega$ the restriction to the $h_{\alpha\beta}^{TT}$ perturbations defines
the consistent truncation \eqref{eq:TTtruncation},  which we use following  \cite{Bachas:2022etu}. Substituting \eqref{eq:TTtruncation} into 
\eqref{eq:SVTvector} yields
$\partial_\zeta H_\zeta=i\omega H_t$,
and 
$\partial_\zeta H_t=i\omega H_\zeta$, whose general solution decomposes into left- and right-moving
components:
\begin{subequations}\label{left right decomp}
\begin{align}\label{Janus hpm}
    e^{i\omega t}H_\zeta(y,\zeta) &=\left( e^{i\omega (t+\zeta)}H_+(y)+e^{i\omega (t-\zeta)}H_-(y)\right),\\
    e^{i\omega t}H_{t}(y,\zeta) &= \left(e^{i\omega (t+\zeta)}H_+(y)-e^{i\omega (t-\zeta)}H_-(y)\right).
\end{align}
\end{subequations}
Substituting \eqref{left right decomp} into the two last Einstein equations \eqref{Janus Einstein pert zz}-\eqref{Janus Einstein pert zt}, we see that \(H_\pm\)  both satisfy 
\begin{equation}
\left(
\partial_y^2
+2\frac{a'(y)}{a(y)}\partial_y
\right)H_\pm(y)
=
\frac{1}{a^2(y)} \partial_y\left(a^2(y)\partial_yH_\pm(y) \right)=0.
\end{equation}
These equations can be solved for $H_{\pm}(y)$, yielding:
\begin{equation}\label{eq:JanusHpm}
     H_\pm(y) =  A_\pm +  B_\pm \int_0^y  \frac{l}{a(\tilde{y})^2}d\tilde{y}     =A_\pm + \dfrac{2B_\pm}{\sqrt{b(2-b)}}\,\operatorname{\arctanh}\!\left[\sqrt{\dfrac{b}{2-b}}\tanh\dfrac{y}{l}\right] \,.
\end{equation}
The corresponding full metric perturbation \eqref{eq:metricpertgenapp} reads: 
\begin{align}\label{eq:TTmetpertG12}
h_{\alpha\beta}^{\mathrm{TT}}(y,\zeta,t)=
\frac{e^{i\omega t}}{l^2} 
\begin{pmatrix}
 H_{\zeta} & 
 H_t \\
 H_t & 
 H_\zeta \\
\end{pmatrix}
=
\frac{1}{l^2} e^{i\omega (t+\zeta)}H_+(y)
\begin{pmatrix}
 1 & 
 1 \\
 1 & 
 1 \\
\end{pmatrix}
+\frac{1}{l^2} e^{i\omega (t-\zeta)} H_-(y)
\begin{pmatrix}
  1 & 
 -1 \\
 -1 & 
 1  \\
\end{pmatrix}.
\end{align}
In what follows, it will be useful to define 
\begin{equation}
\Omega\equiv\frac{\sqrt{b(2-b)}}{2\arctanh \left(\sqrt{\frac{b}{2-b}}\right)}. 
\end{equation}

\subsection{Computation of $\mathcal{T}_L$ following the main text}\label{app:Januspart1}

In this section, we outline the calculation presented in the main text.
We now set $\omega=0$. Since the component functions $H_\zeta$ and
$H_t$ enter the following calculation in parallel, we relabel them
$H_1$ and $H_2$, respectively. 
The metric perturbation can then be
written as
\begin{equation}
h_{\alpha\beta}^{\mathrm{TT}}(y)
=
\frac{1}{l^2}
\begin{pmatrix}
H_1(y) & H_2(y)\\
H_2(y) & H_1(y)
\end{pmatrix},
\qquad
\begin{aligned}
H_1(y)&\equiv H_\zeta(y)=H_+(y)+H_-(y),\\
H_2(y)&\equiv H_t(y)=H_+(y)-H_-(y).
\end{aligned}
\end{equation}
The explicit functions read
\begin{equation}\label{eq:Janus-static-Hi}
H_i(y)
=
A_i
+\frac{2B_i}{\sqrt{b(2-b)}}
\operatorname{arctanh}\!\left[
\sqrt{\frac{b}{2-b}}
\tanh\!\left(\frac{y}{l}\right)
\right],
\quad i=1,2,
\qquad
\begin{aligned}
A_1&\equiv A_++A_-,
&
B_1&\equiv B_++B_-,
\\
A_2&\equiv A_+-A_-,
&
B_2&\equiv B_+-B_-.
\end{aligned}
\end{equation}
The perturbation used in the main text,
\eqref{eq:bmodepert}, corresponds to the off-diagonal component of the parametrization above, with
$\beta(y)=\frac{H_2(y)}{l^2}$.
Here we retain in addition the independent diagonal
transverse-traceless component $H_1(y)$. This allows us to consider a
more general static transverse-traceless metric source. As we  show below in equation \eqref{VEV Janus L/R}, this additional component does not
contribute to the energy flux $\langle T_{tx}\rangle$ on either side of the interface even in the presence of sources. 
The asymptotic values of the functions $H_i(y)$ are, up to a factor of
$l^{-2}$, equal to the corresponding components of the boundary-metric sources
in the AdS$_2$ conformal frame. We therefore define
\begin{equation}\label{Janus sources}
    S_i^L \equiv H_i(-\infty) =  A_i-\frac{B_i}{\Omega} \,, \qquad
    S_i^R \equiv H_i(\infty) = A_i+\frac{B_i}{\Omega}, \qquad i=1,2 \,.
\end{equation}

To determine the stress tensor on the left and right boundaries, we
bring the perturbed metric near each asymptotic boundary,
$y\to\pm\infty$, to Fefferman-Graham form in the flat conformal frame
up to order $\rho^2$:\footnote{By flat conformal frame we mean that the zeroth-order metric is flat, excluding the perturbation.}
\begin{equation}\label{janus FG expansion}
ds_{L/R}^2
=
\frac{l^2}{\rho^2}
\left[
d\rho^2
+
\left(
\eta_{\alpha\beta}
+\epsilon\,\delta g^{L/R}_{(0)\alpha\beta}(x,t)
+\epsilon\,\frac{\rho^2}{l^2}
\delta g^{L/R}_{(2)\alpha\beta}(x,t)
+O(\rho^4,\epsilon^2)
\right)
dx^\alpha dx^\beta
\right],
\end{equation}
where $x^\alpha=(x,t)$. 
To do so, we extend the background coordinate
transformation described around \eqref{eq:FGgeneric} to the perturbed
geometry separately near the two boundaries. 
The order-$\epsilon$ corrections are determined by requiring the metric to take the form
\eqref{janus FG expansion} and are chosen to start at order
$\rho^2$ so as not to change the corresponding boundary sources.\footnote{In the Fefferman-Graham gauge \eqref{janus FG expansion},
$g_{(0)\alpha\beta}\equiv \eta_{\alpha\beta}
+\epsilon\,\delta g_{(0)\alpha\beta}$ is the metric source.
The transformation \eqref{eq: direct coordinate change} can be viewed as a composition of the standard, finite $O(\epsilon^0)$ diffeomorphism used to transform from AdS$_2$ slicing to flat slicing, and a small $O(\epsilon)$ correction. 
An infinitesimal $O(\epsilon)$ bulk diffeomorphism with
$\delta\rho/\rho=-\sigma(x) + O(\rho^2)$ and $\delta x^\alpha = \xi_{(0)}^\alpha + O(\rho^2) $ induces a boundary Weyl transformation and diffeomorphism of the form 
$\delta_{\text{Weyl+diff}}\, g_{(0)}=2\sigma g_{(0)}+\mathcal{L}_{\xi_{(0)}}g_{(0)}$. 
However, in the transformation \eqref{eq: direct coordinate change}, the $O(\epsilon)$ correction has  $\delta\rho=O(\rho^3),\,
\delta x^\alpha=O(\rho^2)$, and thus leaves $g_{(0)}$ unchanged. 
In section \ref{app:Januspart2}, by contrast, the
coordinate transformations contain leading $O(\epsilon\rho^0)$ terms
that implement boundary Weyl and diffeomorphism transformations and
are chosen to set $g_{(0)\alpha\beta}=\eta_{\alpha\beta}$ separately on
each side, which is always possible for $d=2$. 
In CFT terms, the order-$\epsilon$ corrections in the first calculation
keep the metric source fixed, whereas those in the second calculation choose a
sourceless flat conformal frame. 
See Ref.~\cite{Imbimbo:1999bj}, especially Eqs.~(2.2)-(2.9), noting that
their radial coordinate is proportional to our $\rho^2$.\label{fn:FG-source-change}} 
On the right-hand patch, $x>0$, the transformation reads
\begin{subequations}\label{eq: direct coordinate change}
\begin{align}\label{eq: direct coordinate change 1}
\frac{y}{l}
&=
-\log\left(
\frac{\sqrt{1-b}\,\rho}{2x}
\right)
+\frac{\rho^2}{4x^2}
+\epsilon\frac{\rho^2}{l^2}
\,g_{\epsilon}^{(2)}(x)+\dots,
\\
\zeta
&=
\sqrt{x^2+\rho^2}
+\epsilon\frac{\rho^2}{l}f_{\epsilon}^{(2)}(x)+\dots,
\\
t_{\text{old}}& =
t+\epsilon\frac{\rho^2}{l}h_{\epsilon}^{(2)}(x)+\dots.
\end{align}
\end{subequations}
Substituting this transformation into
\eqref{janus perturbations metric}, the functions
$f_\epsilon^{(2)}(x)$ and $h_\epsilon^{(2)}(x)$ are fixed by requiring
the order-$\rho^{-1}$ terms in $g_{\rho x}$ and $g_{\rho t}$ to vanish,
while $g_\epsilon^{(2)}(x)$ removes the order-$\rho^0$ correction to
$g_{\rho\rho}$:
    \begin{align}\label{eq:fghG1}
     g_{\epsilon}^{(2)}(x)
    = -\,\frac{1
    }{4
    }
    S_1^R
    \,, \qquad
    f_{\epsilon}^{(2)}(x)
    &= -\,\frac{x
    }{2l
    } S_1^R
    \,, \qquad
    h_{\epsilon}^{(2)}(x)
    = \,\frac{x
    }{2l
    }S_2^R
    \,.
    \end{align}
Near the left boundary, the analogous Fefferman-Graham map is obtained
from the right-hand map by $y\to-y$, $x\to-x$, and
$S_i^R\to S_i^L$.

The resulting Fefferman--Graham coefficients in the right-hand patch,
$x>0$, are
\begin{equation}
\delta g_{(0)\alpha\beta}^{R}
=
\frac{x^2}{l^2}
\begin{pmatrix}
S_1^R & S_2^R\\
S_2^R & S_1^R
\end{pmatrix},\qquad 
\delta g_{(2)\alpha\beta}^{R}
=
-\frac{1}{2}
\begin{pmatrix}
B_1 & B_2\\
B_2 & B_1
\end{pmatrix}
-\frac{S_1^R}{2}\eta_{\alpha\beta}.
\end{equation}
Near the left boundary, the analogous transformation gives\footnote{The left-hand expressions follow from the right-hand ones
by the reflection described under \eqref{eq:fghG1} together with sending 
$B_i\to-B_i$ (since the function multiplying $B_i$ in
\eqref{eq:Janus-static-Hi} is odd in $y$) and reversing the sign of the $xt$ components (due to reflection in $x$).}
\begin{equation}
\delta g_{(0)\alpha\beta}^{L}=
\frac{x^2}{l^2}
\begin{pmatrix}
S_1^L & -S_2^L\\
-S_2^L & S_1^L
\end{pmatrix},\qquad 
\delta g_{(2)\alpha\beta}^{L}
=
\frac{1}{2}
\begin{pmatrix}
B_1 & -B_2\\
-B_2 & B_1
\end{pmatrix}
-\frac{S_1^L}{2}\eta_{\alpha\beta},
\end{equation}
where the matrices are written in the coordinate order
$x^\alpha=(x,t)$, so that
$\eta_{\alpha\beta}=\operatorname{diag}(1,-1)$. 
We can now extract the VEV of the stress tensor by evaluating \cite{deHaro:2000vlm}:
\begin{equation}\label{eq: stress tensor full expression}
\left\langle T_{\alpha\beta}^{L/R}\right\rangle
=
\frac{  \epsilon}{8\pi l G_N}
\left(
\delta g^{L/R}_{(2)\alpha\beta}
-\eta_{\alpha\beta}
\operatorname{tr}
\left[
\eta^{-1}\cdot
\delta g^{L/R}_{(2)\alpha\beta}
\right]
\right)
+\mathcal O(\epsilon^2).
\end{equation}
This yields
\begin{equation}\label{VEV Janus L/R}
\frac{8\pi lG_N}{\epsilon}
\left\langle T_{\alpha\beta}^R\right\rangle
=
-\frac{1}{2}
\begin{pmatrix}
B_1 & B_2\\
B_2 & B_1
\end{pmatrix}
+\frac{S_1^R}{2}\eta_{\alpha\beta},\qquad
\frac{8\pi lG_N}{\epsilon}
\left\langle T_{\alpha\beta}^L\right\rangle
=
\frac{1}{2}
\begin{pmatrix}
B_1 & -B_2\\
-B_2 & B_1
\end{pmatrix}
+\frac{S_1^L}{2}\eta_{\alpha\beta}.
\end{equation}
The first term in each expression is traceless and depends only on the
normalizable coefficients $B_i$, whereas the second is a local,
source-dependent contribution associated with the $d=2$ Weyl anomaly.
Before imposing any condition on the boundary sources, the $A_i$ and
$B_i$ are independent. Since the stress-tensor VEVs in the AdS$_2$
conformal frame depend only on $B_i$, any explicit dependence on $A_i$
generated upon transforming to the flat frame is anomalous. In the
expressions above, the $A_1$ dependence survives in the diagonal
components through $S_1^{L/R}$. Indeed,
$\left\langle (T^{L/R}){}^{\alpha}{}_{\alpha}\right\rangle
=lR^{L/R}/(16\pi G_N)
=\epsilon S_1^{L/R}/(8\pi lG_N)$, where $R^{L/R}$ is the Ricci scalar of the
boundary metric on the left/right in the zero-order flat conformal frame. By contrast, the anomalous contribution proportional to $A_2$ cancels
from the off-diagonal component, leaving
\begin{equation}
\frac{8\pi lG_N}{\epsilon}
\left\langle T_{tx}^R\right\rangle
=
\frac{8\pi lG_N}{\epsilon}
\left\langle T_{tx}^L\right\rangle
=
-\frac{B_2}{2}.
\end{equation}
This equality holds before setting either boundary source to zero and
expresses conservation of the energy flux across the interface. At this
stage, $A_2$ has therefore dropped out (as it must for the conservation of energy to hold for arbitrary sources on the two sides). 
Consequently,
the transmitted flux may be evaluated on either side, as stated in
footnote~\ref{FN: Ttx evaluation left}, and in particular this shows that we could simply choose to evaluate it on the side with no sources (after imposing the sourceless condition on that side).

To implement the one-sided sourcing prescription of
\eqref{eq:asymsources}, we set the boundary-metric sources on the right
to zero, $\delta g^{R}_{(0)\alpha\beta}=0$. This requires:
\begin{equation}\label{no source right}
    B_1=-A_1 \, \Omega\, ,\qquad B_2=-A_2 \, \Omega \,.
\end{equation}
These relations also imply $S_i^L=2A_i$. In particular, keeping
$A_2$ fixed keeps the left-hand metric source fixed, independently of
$b$, when comparing with the homogeneous theory.
After imposing the conditions \eqref{no source right}, the VEV of the $tx$ component of the stress tensor on the right is:
\begin{equation}
    \frac{8\pi lG_N}{\epsilon} \langle T_{tx} \rangle|_{\text{ICFT}+\mathcal{J}_L} =\frac{1}{2}A_2 \, \Omega\,,
\end{equation}
whereas the same VEV in the presence of the same sources but without the interface, \ie when $b=0$, is:
\begin{equation}
    \frac{8\pi lG_N}{\epsilon} \langle T_{tx} \rangle|_{\text{CFT}_L+\mathcal{J}_L} =\frac{1}{2}A_2 \,.
\end{equation}
We conclude that the transmission coefficient has the expected form:
\begin{equation}
     \mathcal{T}_L= \frac{\langle T_{tx} \rangle |_{\text{ICFT}+\mathcal{J}_L}}{\langle T_{tx} \rangle|_{\text{CFT}_L+\mathcal{J}_L}}=\Omega
      =\frac{\sqrt{b(2-b)}}{2\arctanh \left(\sqrt{\frac{b}{2-b}}\right)}.
\end{equation}
This reproduces the explicit Janus result \eqref{eq:JanusTransmission2dappG} and, through \eqref{eq:JanusTLd2},
agrees with the general formula \eqref{eq:transmission_JanusN}.

\subsection{Computation of $\mathcal{T}_L$ following \cite{Bachas:2022etu}}\label{app:Januspart2}

This second computation is different from the first in two ways.  
First, we keep $\omega\neq0$. Introducing the light-cone coordinates
$x^\pm=t\pm\zeta$ on
each AdS$_2$ slice, the transverse-traceless solution
\eqref{eq:TTmetpertG12} has nonvanishing components
$h_{\pm\pm}(y,\zeta, t)=l^{-2}e^{i\omega x^\pm}H_\pm(y)$, with $H_\pm(y)$ given in
\eqref{eq:JanusHpm}. 
Second, we again bring the metric near each boundary to
Fefferman--Graham form as in \eqref{janus FG expansion}, but now choose
the coordinate transformations to eliminate all sources for the
perturbation, both away from the interface and at the interface itself. In particular, we impose
$\delta g^{L/R}_{(0)\alpha\beta}=0$ on both sides and fix the residual
boundary-coordinate freedom so that the interface remains at $x=0$
with the same time  coordinate parametrization from either side. 
As explained in
footnote~\ref{fn:FG-source-change}, achieving this requires the
coordinate transformations to include leading
$O(\epsilon\rho^0)$ corrections. 
The corrections are chosen to make the boundary metric flat on each side separately, so the
anomalous trace of the stress tensor vanishes.
Equation~\eqref{eq: stress tensor full expression} then reduces to
$\langle T^{L/R}_{\alpha\beta}\rangle
=\epsilon\,\delta g^{L/R}_{(2)\alpha\beta}/(8\pi lG_N)$. 
On the right-hand patch, $x>0$, these coordinate transformations take
the following form: 
\begin{align}
\frac{y}{l}
={}&
-\log\left(
\frac{\sqrt{1-b}\,\rho}{2x}
\right)
+\frac{\rho^2}{4x^2}
+\epsilon\left[
\,g_\epsilon^{(0)}(x,t)
+\frac{\rho^2}{l^2}
\,g_\epsilon^{(2)}(x,t)
\right]+\mathcal O(\epsilon^2,\rho^4),
\\
\zeta
={}&
\sqrt{x^2+\rho^2}
+l \epsilon\left[
f_\epsilon^{(0)}(x,t)
+\frac{\rho^2}{l^2}f_\epsilon^{(2)}(x,t)
\right]
+\mathcal O(\epsilon^2,\rho^4),
\\
t_{\mathrm{old}}
={}&
t+l\epsilon\left[
h_\epsilon^{(0)}(x,t)
+\frac{\rho^2}{l^2}h_\epsilon^{(2)}(x,t)
\right]
+\mathcal O(\epsilon^2, \rho^4).
\end{align}
The functions $f_\epsilon^{(0)}$, $h_\epsilon^{(0)}$, and
$g_\epsilon^{(0)}$ are chosen to cancel the
$O(\epsilon\rho^{-2})$ perturbations of $g_{\alpha\beta}$, so that
$g^R_{(0)\alpha\beta}=\eta_{\alpha\beta}$. 
As in Sec.~\ref{app:Januspart1},
$f_\epsilon^{(2)}$ and $h_\epsilon^{(2)}$ cancel the
$O(\rho^{-1})$ terms in $g_{\rho x}$ and $g_{\rho t}$, while
$g_\epsilon^{(2)}$ removes the $O(\rho^0)$ correction to
$g_{\rho\rho}$. 
We find:
\begin{subequations}\label{eq:myFGg2}
\begin{align}
g_\epsilon^{(0)} &= -\frac{1}{2l^2\omega^3x}\left[
(x\omega-i)\left(A_-+\frac{B_-}{\Omega}\right)e^{i\omega(t-x)}
+(x\omega+i)\left(A_++\frac{B_+}{\Omega}\right)e^{i\omega(t+x)}\right]\notag\\
&\qquad
-l\left(k_-^{R\prime}(t-x)+k_+^{R\prime}(t+x)\right)+\frac{l}{x}\left(-k_-^R(t-x)+k_+^R(t+x)
\right),
\\[2mm]
f_\epsilon^{(0)} &= 
\frac{1}{2l^3\omega^3}\left[-i\left(x^2\omega^2-ix\omega-1\right)\left(A_-+\frac{B_-}{\Omega}\right)e^{i\omega(t-x)}
+i\left(x^2\omega^2+ix\omega-1\right)\left(A_++\frac{B_+}{\Omega}\right)e^{i\omega(t+x)}\right]\notag\\
&\qquad-k_-^R(t-x)+k_+^R(t+x),
 \\[2mm]
h_\epsilon^{(0)} &= 
\frac{x}{2l^3\omega^2}\left[-(1+ix\omega)\left(A_-+\frac{B_-}{\Omega}\right)e^{i\omega(t-x)}
+(1-ix\omega)\left(A_++\frac{B_+}{\Omega}\right)e^{i\omega(t+x)}\right]\notag\\
 &\qquad+k_-^R(t-x)+k_+^R(t+x),
 \\[2mm]
 g_\epsilon^{(2)} &= 
\frac{1}{4\omega^3 x^3}\left[\left(-i+x\omega+ix^2\omega^2\right)\left(A_-+\frac{B_-}{\Omega}\right)e^{i\omega(t-x)}
+\left(i+x\omega-ix^2\omega^2\right)\left(A_++\frac{B_+}{\Omega}\right)e^{i\omega(t+x)}\right]\notag\\
&\qquad+\frac{l^3}{2x^3}\left(x^2\left(k_-^{R\prime\prime}(t-x)-k_+^{R\prime\prime}(t+x)\right)+x\left(k_-^{R\prime}(t-x)+k_+^{R\prime}(t+x)\right)+k_-^R(t-x)-k_+^R(t+x)\right),
\\[2mm]
f_\epsilon^{(2)} &= 
\frac{1}{4l \omega^3x^2}\left[-\left(x^3\omega^3-x\omega+i\right)\left(A_-+\frac{B_-}{\Omega}\right)e^{i\omega(t-x)}
+\left(-x^3\omega^3+x\omega+i\right)\left(A_++\frac{B_+}{\Omega}\right)e^{i\omega(t+x)}\right]\notag\\
&\qquad+\frac{l^2}{2x^2}\left(x^2\left(k_-^{R\prime\prime}(t-x)-k_+^{R\prime\prime}(t+x)\right)+2x\left(k_-^{R\prime}(t-x)+k_+^{R\prime}(t+x)\right)+k_-^R(t-x)-k_+^R(t+x)\right),
 \\[2mm]
h_\epsilon^{(2)} &= 
\frac{x}{4l}\left[-\left(A_-+\frac{B_-}{\Omega}\right)e^{i\omega(t-x)}
+\left(A_++\frac{B_+}{\Omega}\right)e^{i\omega(t+x)}\right]\notag\\
&\qquad+\frac{l^2}{2}\left(k_-^{R\prime\prime}(t-x)+k_+^{R\prime\prime}(t+x)\right).
\end{align}
\end{subequations}
Here $k_\pm^R(t\pm x)$ are arbitrary chiral functions. They parametrize
residual Fefferman-Graham diffeomorphisms whose modes generate the two
copies of the Virasoro algebra. The form of the
left-hand, $x<0$, Fefferman-Graham map is obtained by $y\to-y$ and $x\to-x$.
In the expressions above, this leaves $A_\pm$ unchanged, sends
$B_\pm\to-B_\pm$, and replaces $k_\pm^R$ by independent functions
$k_\mp^L$. The exchange of chiral labels follows from
$t+x\leftrightarrow t-x$ under $x\to-x$. The labels on $A_\pm$
and $B_\pm$ refer to the bulk dependence $t\pm\zeta$ and are thus retained. Since we are considering a perturbation of fixed frequency $\omega$,
we choose the chiral functions to have the same Fourier frequency dependence as the other elements of the calculation:
\begin{equation}
k_\pm^{L/R}(t\pm x)
=
\frac{\bar{k}_\pm^{L/R}}{2(i\omega l)^3}
e^{i\omega(t\pm x)},
\end{equation}
where $\bar{k}_\pm^{L/R}$ are constants.

To display the dependence on the AdS$_2$-frame sources, we define their chiral combinations, cf. equation \eqref{Janus sources}
\begin{equation}
S_\pm^R
\equiv \frac{1}{2}\left(S_1^R\pm S_2^R\right)
= A_\pm+\frac{B_\pm}{\Omega},
\qquad
S_\pm^L
\equiv \frac{1}{2}\left(S_1^L\pm S_2^L\right)
= A_\pm-\frac{B_\pm}{\Omega}.
\end{equation}
The stress-tensor VEVs then read:\footnote{As explained following equation \eqref{eq:myFGg2}, the forms of the
left- and right-hand expressions are related by the reflections
$y\to-y$ and $x\to-x$. The boundary reflection exchanges
$x^+\leftrightarrow x^-$ and hence
$T_{++}\leftrightarrow T_{--}$. Consequently,
$
\left\langle T_{\pm\pm}^R(t,x)\right\rangle
=
\left.
\left\langle T_{\mp\mp}^L(t,-x)\right\rangle
\right|_{
S_\pm^L\to S_\pm^R,\,
B_\pm\to-B_\pm,\,
\bar{k}_\mp^L\to\bar{k}_\pm^R
}.
$
}
\begin{subequations}
\begin{align}
\text{right incoming:}\qquad
\frac{8\pi lG_N}{\epsilon}\left\langle T_{++}^R\right\rangle
&=
\frac{1}{2}e^{i\omega(t+x)}
\left(S_+^R-B_+-\bar{k}_+^R\right),
\\
\text{right  outgoing:}\qquad
\frac{8\pi lG_N}{\epsilon}\left\langle T_{--}^R\right\rangle
&=
\frac{1}{2}e^{i\omega(t-x)}
\left(S_-^R-B_--\bar{k}_-^R\right),
\\
\text{left outgoing:}\qquad
\frac{8\pi lG_N}{\epsilon}\left\langle T_{++}^L\right\rangle
&=
\frac{1}{2}e^{i\omega(t+x)}
\left(S_-^L+B_--\bar{k}_+^L\right),
\\
\text{left incoming:}\qquad
\frac{8\pi lG_N}{\epsilon}\left\langle T_{--}^L\right\rangle
&=
\frac{1}{2}e^{i\omega(t-x)}
\left(S_+^L+B_+-\bar{k}_-^L\right).
\end{align}
\end{subequations}

We impose the following conditions to obtain the transmission
coefficient. As explained in
footnote~\ref{fn:FG-source-change}, the leading terms in the
Fefferman-Graham maps induce boundary diffeomorphisms and accompanying
Weyl transformations. 
We must choose these transformations so as to  select a conformal frame in which the CFT background is the same as in the unperturbed solution, including the interface itself.
At the
boundary, $f_\epsilon^{(0)}$ gives the transverse component of the
induced diffeomorphism, so requiring the interface at $\zeta=0$ to
remain at $x=0$ in both Fefferman-Graham charts, without being displaced, requires
$f_\epsilon^{(0)}$ to vanish there on both sides. The function
$h_\epsilon^{(0)}$ reparametrizes time along the interface, and using
the same time coordinate on both sides requires it to be continuous
there. The first condition reads: 
\begin{equation}
f_\epsilon^{(0)}(0^-,t)
=
f_\epsilon^{(0)}(0^+,t)
=
0,
\qquad i.e.,\qquad
\begin{cases}
\bar{k}_-^R-\bar{k}_+^R=S_-^R-S_+^R,\\[2pt]
\bar{k}_-^L-\bar{k}_+^L=S_+^L-S_-^L.
\end{cases}
\end{equation}
Adding the two equations above yields the energy-flux conservation
condition at the interface 
$\left\langle T_{--}^L(0^-,t)\right\rangle
+\left\langle T_{++}^R(0^+,t)\right\rangle
=
\left\langle T_{--}^R(0^+,t)\right\rangle
+\left\langle T_{++}^L(0^-,t)\right\rangle$. 
Individually, the equations remove the $1/x$ singularity in the
corresponding left- or right-hand Fefferman-Graham map. Substituting
them into the explicit expressions leaves
\begin{equation}
g_{\epsilon,L}^{(0)}(0^-,t)
=
g_{\epsilon,R}^{(0)}(0^+,t)
=
0,
\end{equation}
so the leading shifts of the  coordinate $y$ agree at the
interface (and therefore on the CFT side the Weyl factor matches on the two sides of the interface, see footnote \ref{fn:FG-source-change}). The remaining condition, which fixes a common time
coordinate along the interface, is
\begin{equation}
h_\epsilon^{(0)}(0^-,t)
=
h_\epsilon^{(0)}(0^+,t),
\qquad i.e.,\qquad
\overline{k}_-^L+\overline{k}_+^L
=
\overline{k}_+^R+\overline{k}_-^R .
\end{equation}

We finally impose two scattering conditions. First, following the
prescription of Ref.~\cite{Bachas:2022etu}, we implement the absence
of a mode emerging from the AdS$_2$ Poincaré horizon by requiring the
$++$ component of the traceless part of the extrinsic curvature of the fixed-$y$ slices to
vanish at linear order. This yields
\begin{equation}
    \hat{K}_{++}
    =
    \sqrt{-\tilde{g}}
    \left(
        K_{++}-\frac{1}{2}\tilde{g}_{++}K
    \right)
    =0,
    \qquad i.e.,\qquad B_+=0\,,
\end{equation}
where $\tilde{g}_{\alpha\beta}$ is the induced metric on the fixed-$y$
slices, $\tilde{g}$ is its determinant, $K_{\alpha\beta}$ is their
extrinsic curvature, and $K$ is its trace. Second, to describe a wave
incident only from the left, we require that there is no incoming mode
on the right:
\begin{equation}
    \left\langle T_{++}^R\right\rangle=0,
\qquad i.e.,\qquad
S_+^R-B_+-\bar{k}_+^R=0.
\end{equation}
Note that once we demand this, $\left\langle T_{--}^R\right\rangle$ represents the transmitted mode and $\left\langle T_{++}^L\right\rangle$ the reflected one.
Combining these conditions with the FG-matching conditions
above, we obtain the transmission coefficient%
\footnote{After imposing all the matching and scattering conditions,
the constants $A_\pm$ cancel from all four boundary stress-tensor
components. The remaining dependence on $B_-$ fixes only the overall
normalization of the incoming wave and therefore cancels from the
transmission coefficient.}
\begin{equation}
    \mathcal{T}_L
    =
    \frac{\left\langle T_{--}^R\right\rangle}
         {\left\langle T_{--}^L\right\rangle}
    =
    \frac{-B_-/2}{-B_-/(2\Omega)}
    =
    \Omega
    =
    \frac{\sqrt{b(2-b)}}
    {2\operatorname{arctanh}\!\left[
        \sqrt{\frac{b}{2-b}}
    \right]}.
\end{equation}

\end{document}